\documentclass[reprint,superscriptaddress,prb,aps]{revtex4-2}

\usepackage{mathtools}
\usepackage{amsmath}
\usepackage{amsthm}
\usepackage{amssymb}
\usepackage{graphicx}
\usepackage{bm}
\usepackage{siunitx}
\usepackage[colorlinks,linkcolor=blue, urlcolor=blue,citecolor=blue,bookmarks=false]{hyperref}

\usepackage{xcolor}

\begin{document}

\title{Correlation-driven electronic nematicity in the Dirac semimetal BaNiS$_{2}$}

\author{C. J. Butler}
\email{christopher.butler@riken.jp}
\affiliation{RIKEN Center for Emergent Matter Science, 2-1 Hirosawa, Wako, Saitama 351-0198, Japan}

\author{Y. Kohsaka}
\affiliation{RIKEN Center for Emergent Matter Science, 2-1 Hirosawa, Wako, Saitama 351-0198, Japan}

\author{Y. Yamakawa}
\affiliation{Department of Physics, Nagoya University, Furo-cho, Nagoya 464-8602, Japan}

\author{M. S. Bahramy}
\affiliation{Department of Physics \& Astronomy, University of Manchester, Oxford Rd., Manchester M13 9PL, United Kingdom}

\author{S. Onari}
\affiliation{Department of Physics, Nagoya University, Furo-cho, Nagoya 464-8602, Japan}

\author{H. Kontani}
\affiliation{Department of Physics, Nagoya University, Furo-cho, Nagoya 464-8602, Japan}

\author{T. Hanaguri}
\email{hanaguri@riken.jp}
\affiliation{RIKEN Center for Emergent Matter Science, 2-1 Hirosawa, Wako, Saitama 351-0198, Japan}

\author{S. Shamoto}
\affiliation{Neutron Science and Technology Center, Comprehensive Research Organization for Science and Society, Tokai, Ibaraki 319-1106, Japan}
\affiliation{Department of Physics, National Cheng Kung University, Tainan, Taiwan 70101, Republic of China}

\begin{abstract}

In BaNiS$_{2}$ a Dirac nodal-line band structure exists within a two-dimensional Ni square lattice system, in which significant electronic correlation effects are anticipated.
Using scanning tunneling microscopy, we discover signs of correlated-electron behavior, namely electronic nematicity appearing as a pair of $C_{2}$-symmetry striped patterns in the local density-of-states at $\sim$60~meV above the Fermi energy.
In observations of quasiparticle interference, as well as identifying scattering between Dirac cones, we find that the striped patterns in real space stem from a lifting of degeneracy among electron pockets at the Brillouin zone boundary.
We infer a momentum-dependent energy shift with $d$-form factor, which we model numerically within a density wave equation framework that considers spin-fluctuation-driven nematicity. 
This suggests an unusual mechanism driving the nematic instability, stemming from only a small perturbation to the Fermi surface,  in a system with very low density of states at the Fermi energy.
The Dirac points lie at nodes of the $d$-form factor, and are almost unaffected by it.
These results highlight BaNiS$_{2}$ as a unique material in which Dirac electrons and symmetry-breaking electronic correlations coexist.

\end{abstract}

\maketitle

Materials in which Dirac fermions and significant electronic correlations coexist represent an intersection of two active frontiers of condensed matter research and might host a range of useful or interesting new phenomena \cite{Keimer2017,Li2010, Cao2018}. Correlations between electrons can result in instability towards a number of spontaneous symmetry-breaking configurations including magnetic order, charge density waves and unconventional superconductivity. They can also lead to nematic order, which plays an important role in the cuprate and Fe-based superconductors \cite{Lawler2010, Fujita2014, Chuang2010, Fernandes2014}. Basic questions include whether topological band structures can survive in the presence of significant electronic correlations, and whether, or in what way, the topological classification of a system can be altered by correlation-driven symmetry-breaking effects such as nematicity, in which translational symmetries are preserved while rotational symmetry is broken \cite{Kivelson1998, Oganesyan2001, Fradkin2010}. However, the impact of such symmetry breaking effects on Dirac fermions remains to be fully elucidated.

Correlations can be introduced through the inclusion of transition metals in the material composition, while otherwise ensuring the requisite crystal symmetries for an interesting topological phase \cite{Schoop2018}. Such a situation is realized in the BaCo$_{1-x}$Ni$_{x}$S$_{2}$ series of compounds. It has recently been shown that the end member of the series, BaNiS$_{2}$, hosts four simple Dirac nodal-lines that are well-isolated from other bands and traverse the Brillouin zone (BZ) with fairly weak dispersion along the layer-perpendicular axis \cite{SantosCottin2021, Nilforoushan2021}. These properties make it one of the systems closest to the ideal of a two-dimensional (2-D) Dirac fermion system. The bands emanating from these nodal-lines have been shown to be susceptible to renormalization under ultrafast light pulses \cite{Nilforoushan2020}, and to a controlled shifting of the Dirac node energy and wavevector upon Co substitution \cite{Nilforoushan2021}. 

\begin{figure*}
\centering
\includegraphics[scale=1]{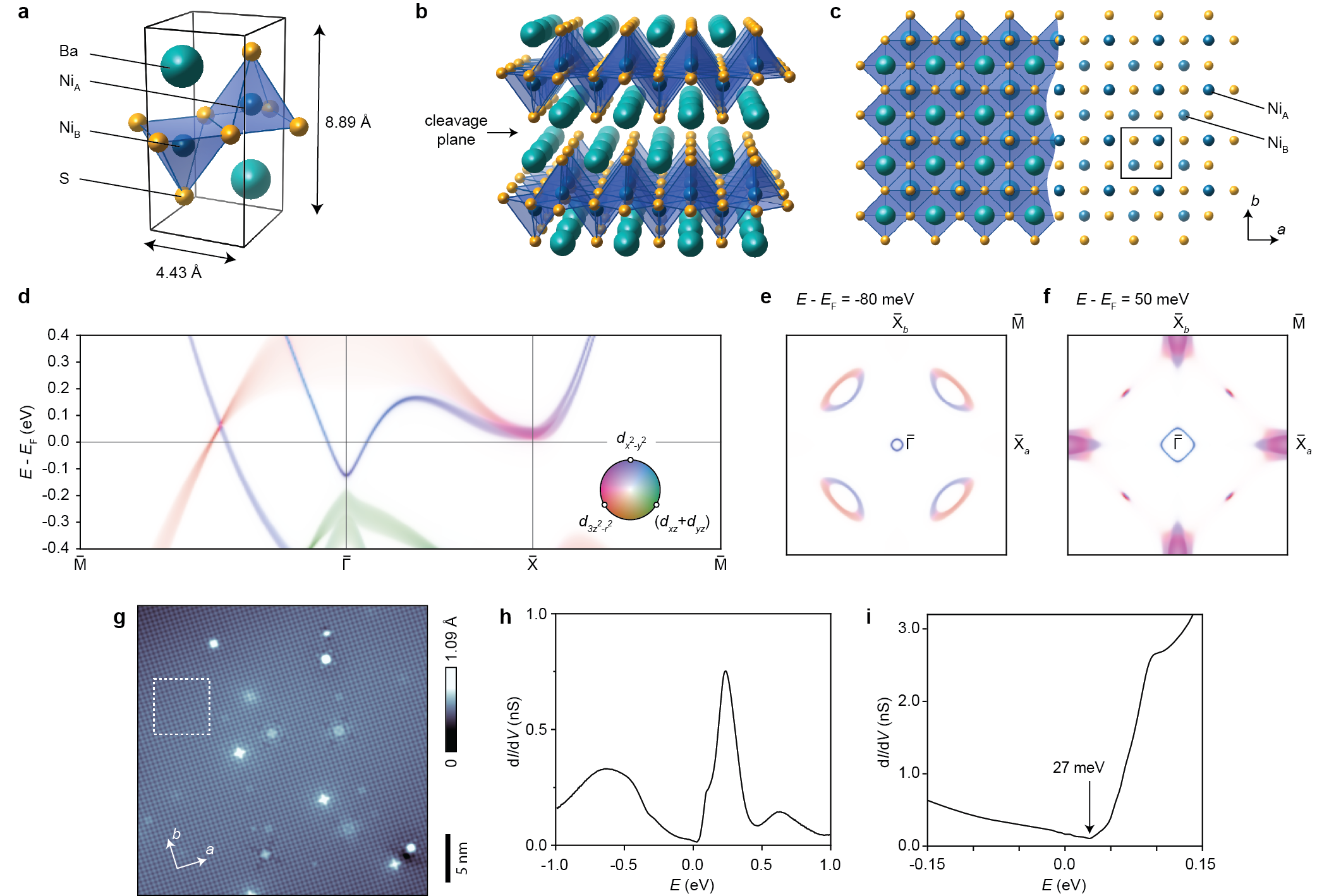}
\caption{
\label{fig:1} \textbf{Overview of the crystal and low-energy electronic structures of BaNiS$_{2}$.}
(a) A depiction of the primitive cell \cite{Grey1970} and (b) the quasi-two-dimensional crystal structure, showing the location of the cleavage plane. (c) A top-down view of the surface lattice. The cutaway view shows the Ni$_{\mathrm{A}}$ and Ni$_{\mathrm{B}}$ square nets. The primitive cell is shown with a black square. Unit-cell and lattice images were created using VESTA \cite{VESTA}. (d) Calculated surface band structure shown along high-symmetry lines. The orbital composition for four of the Ni $d$ orbitals is indicated according to the color key, with the radial axis (lightness) signifying the spectral weight summing over all orbitals. The $d_{xy}$ orbital is not shown as its spectral weight is negligible in this energy range. (e) A constant-energy cross-section through $A(\textbf{k}, E)$ at $E$ = -80~meV, and (f) a cross-section at $E$ = 50~meV. In these calculations the $P4/nmm$ space group is assumed, resulting in $C_{4}$ symmetry of the constant-energy cross-sections, but this assumption will be overturned by the observations of nematicity shown below. In anticipation of this, we draw a distinction between the points $\overline{\mathrm{X}}_{a}$ and $\overline{\mathrm{X}}_{b}$ at the zone boundary. (g) A typical constant-current STM topograph (setpoints $V$ = 0.1~V, $I$ = 100~pA). The observed atomic corrugations likely correspond to the uppermost Ni$_{\textrm{A}}$ square net. (h) A $\frac{\textrm{d}I}{\textrm{d}V}(E)$ conductance curve after averaging over the field of view marked as a white dashed square in (g). (i) A $\frac{\textrm{d}I}{\textrm{d}V}(E)$ curve averaged over the same area, focusing on the region nearer to the Fermi energy $E_{\textrm{F}}$.}
\end{figure*}

The $d$ orbitals of the transition metal square net in BaCo$_{1-x}$Ni$_{x}$S$_{2}$ introduce correlated-electron behavior, most obviously magnetism, but also further interesting properties. While BaNiS$_{2}$ is a paramagnetic metal, BaCoS$_{2}$ is an antiferromagnetic insulator \cite{Mandrus1997}. In the BaCo$_{1-x}$Ni$_{x}$S$_{2}$ solid solution, at $x$ = 0.22 a metal-insulator transition occurs, for which both purely electronic \cite{Krishnakumar2001, Sato2001, Guguchia2019} and structural mechanisms \cite{Schueller2020} have been proposed. Although superconductivity is absent, the temperature-composition phase diagram otherwise resembles that of the layered cuprates, and the above properties make BaCo$_{1-x}$Ni$_{x}$S$_{2}$ an attractive system for the exploration of possible composition-tuned Mott-Hubbard physics on a square lattice, and the search for correlation-driven symmetry breaking phenomena. Even the metallic state of pure BaNiS$_{2}$, furthest from the metal-insulator transition, has been shown to be somewhat anomalous, with $T$-linear resistivity below $T$ = 2~K possibly hinting at the presence in the phase diagram of a quantum critical point \cite{SantosCottin2016a}.

In this work we use scanning tunneling microscopy (STM) to explore the electronic structure at the cleaved surface of BaNiS$_{2}$, observing both nearly ideal topological Dirac fermionic behavior and clear symmetry breaking in the form of bond-order nematicity within the surface Ni square net. In a narrow energy range approximately 60~meV above $E_{\mathrm{F}}$, we observe a lowering of symmetry in the surface local density-of-states, from the expected $C_{4}$ symmetry to two perpendicular striped patterns, each of $C_{2}$ symmetry, and separated by an energy of $\sim$12~meV. From a corresponding lowering of the symmetry in quasiparticle interference (QPI) patterns, we infer a momentum-dependent energy shift of $d$-form factor. The Dirac points are located at the nodes of this form factor, and are therefore left unaltered. Finally we suggest a mechanism, modeled in the density wave equation framework, of interfering spin fluctuations as the driver of the nematic order.

\section*{Properties of the $\mathrm{BaNiS_{2}(001)}$ surface}

BaNiS$_{2}$ is a layered quasi-2-D material with a structure belonging to the nonsymmorphic space group $P4/nmm$, with lattice parameters reported as $a$ = $b$ = 4.43~\AA\ and $c$ = 8.89~\AA\ \cite{Grey1970}. Its structure is composed of NiS$_{5}$ pyramids whose orientation alternates about the basal plane, and whose basal edges are shared with their neighbors, as depicted in Fig. 1(a). Sandwiching these layers are Ba atoms located between, and almost in the plane of, the apical S of the NiS$_{5}$ pyramids. As indicated in Fig. 1(b), cleavage can occur between adjacent BaS layers revealing (001) facets. A top-down view of the surface lattice is shown in Fig. 1(c). The cutaway on the right-hand side of the image shows the Ni$_{\mathrm{A}}$ and Ni$_{\mathrm{B}}$ lattices.

Figure 1(d) shows the (001) surface spectral function $A(\mathbf{k}, E)$ along chosen high-symmetry axes of the BZ. 
The orbital composition of the bands are shown using hue as indicated in the inset.
Approximately halfway along the $\overline{\Gamma \mathrm{M}}$ line, there exists a Dirac cone. This feature represents the surface projection of a 3-D cone emanating from one of the four Dirac nodal-lines running between the $k_{\textrm{z}} = \frac{\pi}{c}$ and $- \frac{\pi}{c}$ planes \cite{Nilforoushan2021}. The electron-like pocket surrounding the $\overline{\Gamma}$-point results from a feature previously described as a pinched electron-like tube connecting the $k_{\textrm{z}} = \pm \frac{\pi}{c}$ faces of the BZ \cite{Klein2018}. A third noteworthy feature is the electron-like band centered around the $\overline{\mathrm{X}}$ point. This results from the collapse into two dimensions of a band surrounding each of the R points at the edges of the bulk BZ, and the previously described Rashba splitting within these bands is also observed here \cite{SantosCottin2016b, Slawinska2016}.

In Fig. 1(e) we show a constant-energy cross-section taken from $A(\mathbf{k}, E)$ at $E$ = -80~meV. Here the electron-like pocket around the bulk $\Gamma$ point, and hole-pockets stemming from the Dirac nodal-lines, are seen to collapse into a circle and a set of ellipses, respectively, in the surface BZ. The Dirac cones are composed of a hybridization of the Ni $d_{x^{2}-y^{2}}$ and $d_{z^{2}}$ orbitals. Below the Dirac points, these orbitals contribute the inner ($\overline{\Gamma}$-facing) and outer ($\overline{\mathrm{M}}$-facing) arcs, respectively \cite{Nilforoushan2021}, and this is reversed above the Dirac points as seen at the left-hand-side of Fig. 1(d). Fig. 1(f) shows a constant-energy cross-section at $E$ = 50~meV, an energy which coincides closely with both the Dirac nodes and the bottom of the electron bands around the $\overline{ \mathrm{X}}$ point. Like the Dirac cones, these bands are composed of a hybridization of $d_{x^{2}-y^{2}}$ and $d_{z^{2}}$ orbitals. We point out that these numerical calculations do not accurately reflect the energies of these features, which are to be properly established through observations presented below (and to some extent in the Supplemental Information).

A typical constant-current STM topography image acquired at a cleaved surface is shown in Fig. 1(g). The observed atomic corrugations almost certainly represent the Ni square net, and specifically Ni$_{\mathrm{A}}$. (The periodicity of the corrugations corresponds to one observed atom per surface unit cell -- see Supplemental Information.)

Figures 1(h) and 1(i) show typical tunneling conductance curves, denoted by $\frac{\textrm{d}I}{\textrm{d}V}(E)$, where by convention $E = eV$ ($e$ being the electron charge and $V$ the sample bias). These are averaged over the $6\times6$~nm$^{2}$ field of view marked as a white dashed square in Fig. 1(g). The curves indicate a density of states close to an ideal semimetal, i.e. becoming small near the Fermi energy. In Fig. 1(i), the `v'-shaped minimum at $E \approx$ 27~meV can be taken to indicate the approximate energy of the Dirac crossings. The shallow and roughly linear onset toward lower energy likely corresponds to Dirac cones emerging from these crossings. Additional spectroscopy measurements acquired under magnetic fields are shown in the Supplementary Information. These indicate the emergence of a field-independent Landau level, indicating a Berry's phase of $\pi$ associated with each Dirac point, as well as giving a more precise estimate of the Dirac points' energy. We note that $E_{\mathrm{F}}$ is located near the Dirac point energy, where the density-of-states is minimal. This situation would generally be unfavorable for an electronic instability. However, as we will show below, BaNiS$_{2}$ exhibits a likely correlation-driven electronic nematicity.

\section*{Observation of nematic order}

\begin{figure*}
\centering
\includegraphics[scale=1]{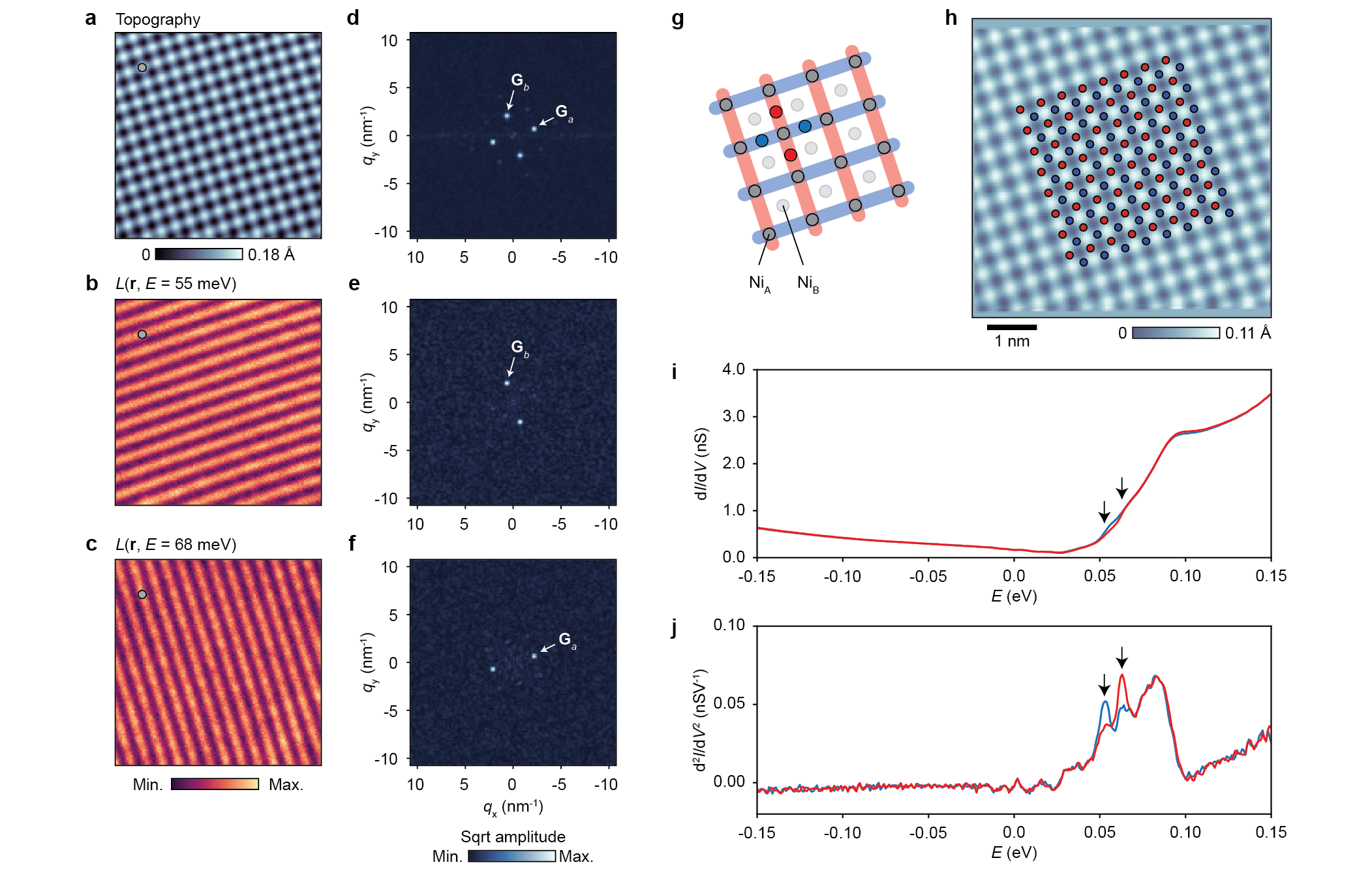}
\caption{\label{fig:2} \textbf{High-resolution spectroscopic imaging.}
(a) Constant-current topography in the $6\times6$~nm$^{2}$ field of view marked by the dashed square in Fig. 1(g) (setpoints $V$ = 0.1~V, $I$ = 500~pA). (b,c) $L$ images at selected energies ($V_{\textrm{mod}}$ = 1~mV). The grey dots in (a--c) mark a single Ni$_{\mathrm{A}}$ site, and show that the lines of high intensity in the $L$ images coincide exactly with the Ni$_{\mathrm{A}}$ lattice. (d--f) Fourier transformations of the images in (a--c). (g) A depiction of the Ni$_{\mathrm{A}}$ square net and its two bond-centered sublattices, marked by arrays of blue and red dots. (h) Sampling of $\frac{\textrm{d}I}{\textrm{d}V}(E)$ data in the same field-of-view as in (a) (setpoints $V$ = 0.15~V, $I$ = 300~pA, and $V_{\textrm{mod}}$ = 2.5~mV), on a $8\times8$ grid on each bond-centered sublattice. (i) $\frac{\textrm{d}I}{\textrm{d}V}(E)$ curves after averaging over each of the sets of sampling points. Each arrow marks a subtle kink in the respective curve. (j) Derivatives $\frac{\textrm{d}^{2}I}{\textrm{d}V^{2}}(E)$ enhancing the small variations of the curves shown in (i). The splitting between the peaks is $\approx$12~meV.}
\end{figure*}

Figure 2 shows high-resolution spectroscopic imaging that reveals an unexpected symmetry breaking within the local density-of-states. Figure 2(a) shows STM topography over the same $6\times6$~nm$^{2}$ field of view as marked by a white dashed square in Fig. 1(g). A gray dot in the upper left-hand corner marks one of the Ni$_{\mathrm{A}}$ sites for a point of reference. Differential conductance $\frac{\textrm{d}I}{\textrm{d}V}(\mathbf{r}, E)$ was acquired by measuring a $\frac{\textrm{d}I}{\textrm{d}V}(E)$ curve at each pixel. Figures 2(b) and 2(c) show images extracted from the normalized differential conductance defined as $L(\mathbf{r}, E) \equiv \frac{\textrm{d}I}{\textrm{d}V}(\mathbf{r}, E) / \frac{I(\mathbf{r}, E)}{V}$, at $E$ = 55~meV and 68~meV, respectively. The normalization mitigates the so-called `set-point effect' \cite{Kohsaka2007}, an unwanted influence of variations in tip-sample spacing on $\frac{\textrm{d}I}{\textrm{d}V}(\mathbf{r})$ images. The gray dots in Figs 2(b) and 2(c) are carried over from the one in Fig. 2(a). The Fourier transform images of the topographic and $L(\mathbf{r})$ images are shown in Figs. 2(d--f). In Fig. 2(d) the peaks corresponding to the reciprocal lattice vectors $\mathbf{G}_{a}$ and $\mathbf{G}_{b}$ are seen, and show the expected $C_{4}$ symmetry. In Figs. 2(b) and 2(c), we see the two $C_{2}$ symmetry stripe patterns, each commensurate with, and in phase with, one of the surface lattice corrugations [i.e. for each pattern $\mathbf{q}_{a,b} = \mathbf{G}_{a,b}$ as seen in Figs 2(e) and (f)]. 

The peaks in intensity of each striped pattern correspond directly with the rows or columns of the Ni$_{\mathrm{A}}$ lattice, and each stripe lies along a Ni$_{\mathrm{A}}$--Ni$_{\mathrm{A}}$ bond chain (or more strictly, a Ni$_{\mathrm{A}}$--S--Ni$_{\mathrm{A}}$ bond chain). Figure 2(g) shows a cartoon of this situation, in which the blue (red) stripes indicate the pattern at the lower (upper) energy. As well as the in-plane projections of the Ni$_{\mathrm{A}}$ and Ni$_{\mathrm{B}}$ sites, the sites belonging to the two inequivalent bond-centered lattices are also shown, with red and blue dots. As shown in Fig. 2(h) we perform a sampling of the raw conductance data in the same field of view as Figs. 2(a--c), on a 8$\times$8 grid on each of the sublattices. Figure 2(i) shows the resulting $\frac{\textrm{d}I}{\textrm{d}V}(E)$ curves (averaged over the 64 sites of each grid). In each $\frac{\textrm{d}I}{\textrm{d}V}(E)$ curve, a subtle shoulder in the steep onset of states above the conductance minimum is marked with a black arrow. A derivative with respect to energy, shown in Fig. 2(j) helps to clarify the difference between the two curves. A peak in the second-derivative of an $I-V$ curve is generally known to correspond to the availability of an additional tunneling channel,  such as a band edge that becomes available upon reaching the necessary bias. The energy difference between the peaks in the two curves is $\sim$12~meV. 

A question that reasonably follows from the above observation is whether a domain structure of the nematic order exists. Throughout this work, examining patterns of nematic order in two samples and searching over fields-of-view up to $\sim$500~nm, no domain boundaries were observed, and the order was seen to be preserved across a step-terrace morphology including three atomic terraces (see Supplemental Information).

\section*{Manifestation of nematicity in QPI}

The mechanism underlying the observed nematic order might be elucidated by examining the momentum-space electronic structure at the BaNiS$_{2}$ surface, and indirect insights into this can be obtained through interpretations of QPI phenomena. Such measurements can complement previous angle-resolved photo-emission spectroscopy reports \cite{SantosCottin2016b, Nilforoushan2020, Nilforoushan2021}, but can additionally probe unoccupied bands.

\begin{figure*}
\centering
\includegraphics[scale=1]{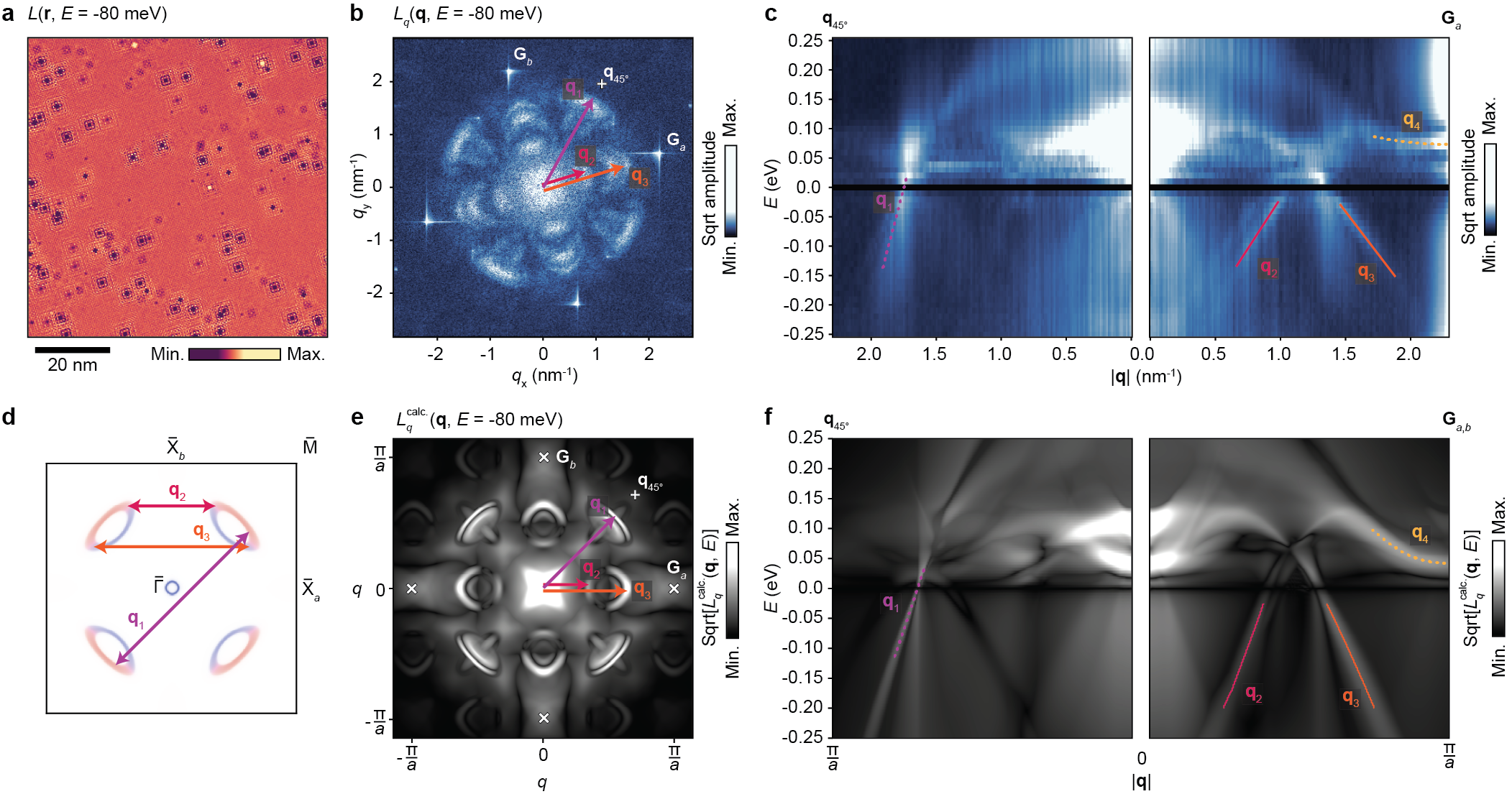}
\caption{\label{fig:3} \textbf{Overview of QPI observations.} 
(a) A normalized conductance image $L(\mathbf{r})$ acquired at $E$ = -80~meV in a $80\times80$~nm$^{2}$ field of view, showing QPI modulations around point defects ($448\times448$ pixels, setpoints $V$ = 0.25~V, $I$ = 1~nA, and $V_{\textrm{mod}}$ = 10~mV). (b) A Fourier transform of the \textit{L} map, labeling the reciprocal lattice vectors $\mathbf{G}_{a}$ and $\mathbf{G}_{b}$. The wavevectors $\mathbf{q}_{1,2,3}$ of QPI modulations discussed in the text, are marked purple, red and orange arrows.  The point $\mathbf{q}_{45^{\circ}}$ marks the endpoint of a vector formed by rotating $\mathbf{G}_{a}$ by $45^{\circ}$. (c) Linecuts through the $L_{q}(\mathbf{q}, E)$ data, from the origin to $\mathbf{G}_{a}$ and also to  $\mathbf{q}_{45^{\circ}}$. The intensity is averaged over the line width of 10 pixels. Here we additionally highlight the wavevector $\mathbf{q}_{4}$, which will be discussed below. The dashed lines for $\mathbf{q}_{1}$ and $\mathbf{q}_{4}$ are guides to the eye.  The solid lines for $\mathbf{q}_{2}$ and $\mathbf{q}_{3}$ are obtained by a fitting procedure described in detail in the Supplementary Information. (d) A schematic of the surface BZ and calculated spectral function at $E$ = -80~meV, showing the likely scattering transitions giving rise to $\mathbf{q}_{1,2,3}$. (e) Calculated QPI pattern $L_{q}^{\mathrm{calc.}}(\mathbf{q})$ at $E$ = -80~meV, based on the spectral function and \textit{T}-matrix formalism (see Methods). Each white `$\times$' symbol marks the position of a reciprocal lattice point. (f) Profiles through $L_{q}^{\mathrm{calc.}}(\mathbf{q}, E)$ highlighting the predicted signals $q^{\textrm{calc.}}_{1,2,3,4}$. The solid red lines mark the ridges in $L_{q}^{\mathrm{calc.}}(\mathbf{q}, E)$, corresponding to $\mathbf{q}_{2}$ and $\mathbf{q}_{3}$, for comparison with the solid lines in (c).}
\end{figure*}

Figure 3(a) shows a selected image from $L(\mathbf{r}, E)$ data acquired in a larger field-of-view, at $E = -80~\mathrm{meV}$, in which modulations resulting from interference caused by scattering of quasiparticles from point defects are seen. Figure 3(b) shows the Fourier transform $\mathcal{F} [L(\mathbf{r}, E = -80~\mathrm{meV})]$. We hereafter refer to such Fourier transformed conductance data loosely as $L_{q}(\mathbf{q}, E)$. The reciprocal lattice vectors are again denoted by $\mathbf{G}_{a}$ and $\mathbf{G}_{b}$. At this energy we identify three prominent scattering vectors, labelled $\mathbf{q}_{1,2,3}$. Figure 3(c) shows linecuts through $L_{q}(\mathbf{q}, E)$, running from the origin to $\mathbf{G}_{a}$, and along a line of the same length running halfway between $\mathbf{G}_{a}$ and $\mathbf{G}_{b}$. The most interesting features are seen in the linecut running from the origin to $\mathbf{G}_{a}$. The branches $\mathbf{q}_{2}$ and $\mathbf{q}_{3}$ form an inverted `v' below $E_{\mathrm{F}}$, and it can be inferred that they represent the shortest and longest possible scattering vectors, respectively, connecting two Dirac cones in adjacent quadrants of the BZ. At positive energy, a branch with apparently very weak dispersion, $\mathbf{q}_{4}$, extends to meet $\mathbf{G}_{a}$. This branch will be discussed in greater detail below.

Figure 3(d) shows the likely origin of each of the scattering signals found at $E$ = -80~meV. We attribute $\mathbf{q}_{1}$ to scattering between Dirac cones on opposite sides of the BZ. We attribute $\mathbf{q}_{2}$ and $\mathbf{q}_{3}$ to scattering between cones in adjacent quadrants of the BZ, and specifically between the least- and most-distance ends of the respective ellipsoids as illustrated in Fig. 3(d). Figure 3(e) and 3(f), show the corresponding calculated QPI pattern and linecuts. All of the labeled QPI branches are qualitatively reproduced.

Comparing the data shown in Figs. 3(c) and (f), we find that the observed dispersions for $\mathbf{q}_{2}$ and $\mathbf{q}_{3}$ are only a factor of 0.58 as high as the calculated ones. With reference to recent related work, we speculate that the reason could be a failure to fully account for electronic correlations that might renormalize the band dispersion at low energies \cite{Fujioka2019}. Therefore the inferred `squeezing' of the band structure near $E_{\mathrm{F}}$ may reflect significant electronic correlation effects in the real material.

A numerical fitting to the $\mathbf{q}_{2}$ and $\mathbf{q}_{3}$ branches, shown as red and orange solid lines in Fig. 3(c), yields apparent velocities $v_{\mathbf{q}_{2}}$ = 8.15$\times10^{4}$~m\,s$^{-1}$ and $v_{\mathbf{q}_{3}}$ = -6.91$\times10^{4}$~m\,s$^{-1}$, respectively. However, these values are related only indirectly to any actual band velocities, since a Fourier transform of QPI modulations visualized the momentum-\textit{transfer} ($q$) space rather than the momentum ($k$) space. By making a comparison between the observed and calculated features in $q$-space, and also by making a $k$-space-to-$q$-space comparison of the calculation results, we can eventually estimate some key properties of the Dirac cones. (See the Supplemental Information for full arguments.) In this way, the band velocities of the inner ($\overline{\Gamma}$-facing) and outer ($\overline{\mathrm{M}}$-facing) sides of each Dirac cone are estimated to be $v_{k,{\mathrm{inner}}}$ = 3.09$\times$10$^{5}$~m\,s$^{-1}$ and $v_{k,{\mathrm{outer}}}$ = -2.98$\times$10$^{5}$~m\,s$^{-1}$, respectively. This is reasonably consistent with a rough estimate drawn from recently reported angle-resolved photoemission spectroscopy results, of $v_{k,{\mathrm{inner}}} \sim v_{k,{\mathrm{outer}}} \approx 2\times10^{5}$~m\,s$^{-1}$ \cite{Nilforoushan2021}.

\begin{figure}
\centering
\includegraphics[scale=1]{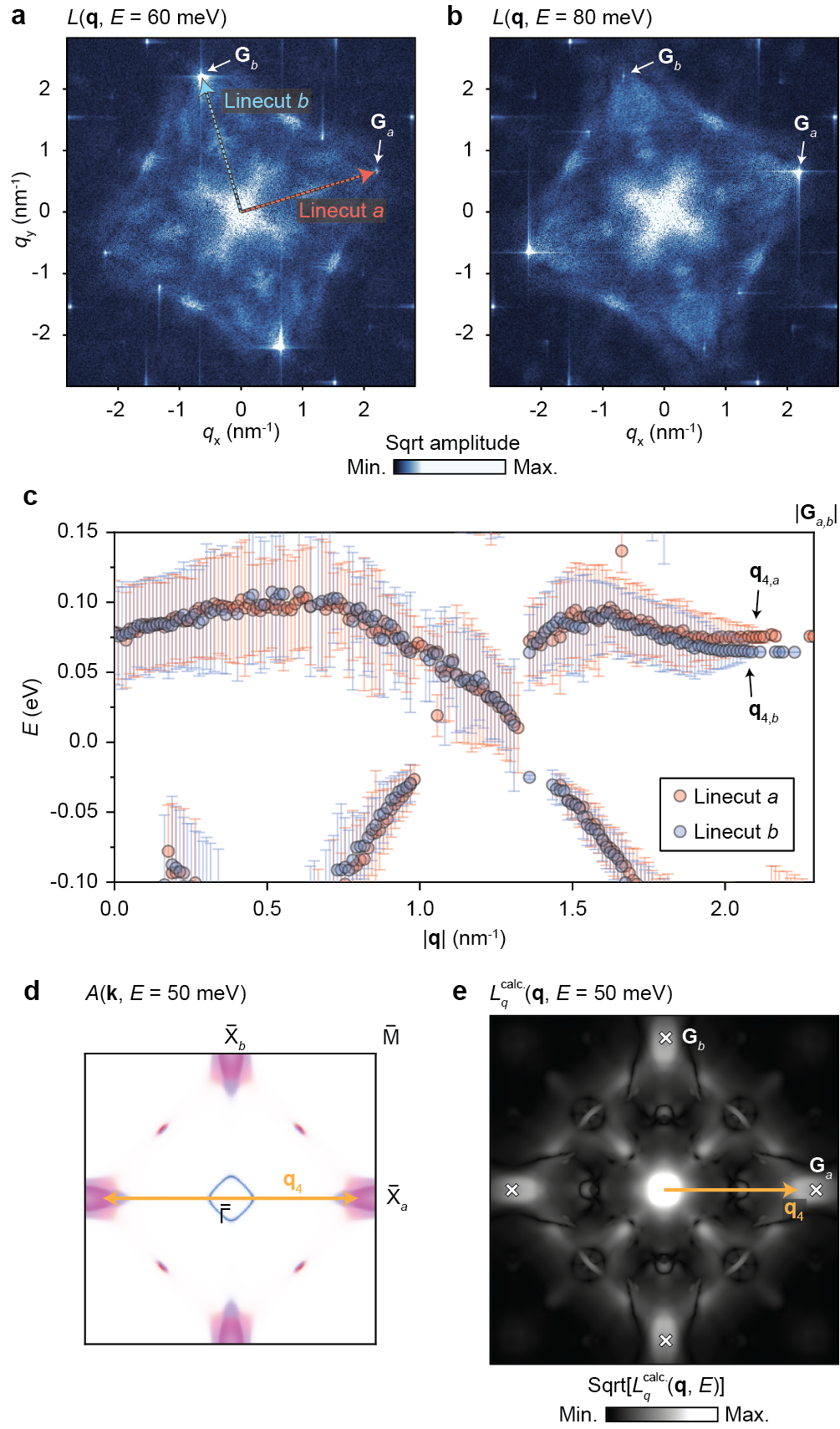}
\caption{\label{fig:4} \textbf{Symmetry breaking in QPI patterns.}
(a) The image $L_{q}(\mathbf{q}, E=\mathrm{60 meV})$, from the same measurement described for Fig. 3 above. The pairs of reciprocal lattice peaks $\mathbf{G}_{a}$ and $\mathbf{G}_{b}$ show a large difference in intensity. (b) The corresponding image at $E$ = 80~meV. Here the alternate set of reciprocal lattice peaks become more intense. (c) Comparison of fitting to $L_{q}(q, E)$ along the $\mathbf{G}_{a}$ and $\mathbf{G}_{b}$ vectors [i.e., along the red and blue dashed lines marked in (a)]. In the region in which $q$ approaches $| \mathbf{G}_{a,b} |$ the branches of points diverge, towards a splitting of $\sim$12~meV at the endpoint. (d) A schematic of the surface BZ and calculated spectral function at $E$ = 50~meV, with the yellow arrow illustrating the scattering vector $q_{4}$. (e) The corresponding calculated QPI pattern $L_{q}^{\mathrm{calc.}}(\mathbf{q})$ at $E$ = 50~meV. See also Fig. 3(f).}
\end{figure}

Having identified the $q$-space signatures of the Dirac cones, we turn our attention to the manifestation in QPI phenomena of the nematic behavior described above. Examining the $L_{q}(\mathbf{q}, E)$ data further, we find that in a small energy range a few tens of meV above $E_{\mathrm{F}}$ there appears a lowering of symmetry similar to that shown in Fig. 2. Two $L_{q}(\mathbf{q})$ images selected from the same data as shown in Fig. 3, at $E$ = 60~meV and $E$ = 80~meV, are shown in Figs. 4(a) and 4(b) respectively. Each image shows an asymmetry in intensity between the pairs of reciprocal lattice peaks $\mathbf{G}_{a}$ and $\mathbf{G}_{b}$ [recall Figs. 2(e) and 2(f)].

Figure 4(c) shows the result of fitting a sum of multiple Lorentzian functions to the $L_{q}(E)|_{q}$ curve at each $q$-point along the lines from $\mathbf{q} = 0$ to each of the reciprocal points $\mathbf{G}_{a}$ and $\mathbf{G}_{b}$. [This is the precursor step to obtaining the linear fits shown in Fig. 3(c), and the full details are described in the Supplemental Information]. The results from each line profile are superimposed along the same axis for comparison. We see that nearly everywhere the peak energies show only small differences, but that a noticeable difference appears as we approach the endpoints $q = \left| \mathbf{G}_{a,b} \right|$. We therefore distinguish between the branches along the two axes by relabeling them as $\mathbf{q}_{4,a}$ and $\mathbf{q}_{4,b}$. As $q$ approaches $|\mathbf{G_{a,b}}|$, the energy splitting between the $\mathbf{q}_{4,a}$ and $\mathbf{q}_{4,b}$ branches approaches $\sim$12~meV. This is consistent with the splitting observed in intra-unit-cell sampling of $\frac{\textrm{d}I}{\textrm{d}V}(\mathbf{r}, E)$ data shown in Fig. 2. The $\sim$10~meV difference in the absolute energies of the two branches as compared to those of the two peaks in Fig. 2(j) is likely due to the fact that the measurements were acquired on two different sample surfaces.

Scattering vectors equal to $\mathbf{G}_{a}$ ($\mathbf{G}_{b}$) in $q$-space serve to connect $\overline{\mathrm{X}}_{a}$ ($\overline{\mathrm{X}}_{b}$) and its counterpart at the opposite edge of the BZ. This suggests that the $\mathbf{q}_{4}$ branch captured in Figs 3 and 4 results from scattering between the electron pockets around the $\overline{\mathrm{X}}_{a}$.Figures 4(d) and 4(e) depict this scenario with reference to the calculated spectral function, and show the appearance of $\mathbf{q}_{4}$ in the calculated QPI pattern.

\section*{Origin of nematicity \textit{via} the density wave equation}

The results presented above describe symmetry-breaking both in the intra-unit cell density-of-states and also in quasiparticle scattering between the pairs of electron-like pockets around opposing $\overline{\mathrm{X}}_{a}$ and $\overline{\mathrm{X}}_{b}$ points. We interpret these results as different manifestations of the same underlying phenomenon. The latter manifestation strongly suggests a $k$-dependent shift of the band energies, that has an opposite sign near $\overline{\mathrm{X}}_{a}$ as compared to $\overline{\mathrm{X}}_{b}$, and this would also give an intuitive explanation for the former manifestation: The electron bands flatten out exactly at the $\overline{\mathrm{X}}_{a,b}$ points so that the last remaining scattering vectors connecting the pairs of bands at the band bottoms are exactly $\mathbf{q}_{a,b} = \mathbf{G}_{a,b}$. The $\mathbf{q}_{a}$ and $\mathbf{q}_{b}$ striped patterns each appear at the energy that the respective band bottoms out. The simplest $k$-dependent energy correction that explains the above results is one with $d$-form factor.

A structural distortion is unlikely as the origin of the observed lifting of degeneracy because BaNiS$_{2}$ is tetragonal at room temperature, and no phase transition is reported down to \textit{T} = 2~K \cite{SantosCottin2016a}. It may be more reasonable to assume an electronic origin. We exclude a simple orbital ordering however, because there is apparently no degeneracy of the relevant orbitals that make up the pocket around the $\overline{\mathrm{X}}$ points (as among $d_{xz}$ and $d_{yz}$, for example). Moreover, the small density-of-states near $E_{\mathrm{F}}$ likely excludes mechanisms stemming from a Fermi-surface-instability. 

For these reasons we adopt a framework based on the density wave (DW) equation, which has previously been used to discuss $d$-form factor energy corrections emerging from many-body correlations in the context of the unconventional superconductors \cite{Onari2012, Onari2016, Kawaguchi2017, Onari2019, Ishida2020, Onari2020, Tazai2022b} as well as exotic density-wave states in other systems \cite{Tazai2022a}. In this scenario, broken symmetry in the charge sector may arise as a many-body correction generated by the interference of spin fluctuations.

\begin{figure}[t]
\centering
\includegraphics[scale=1]{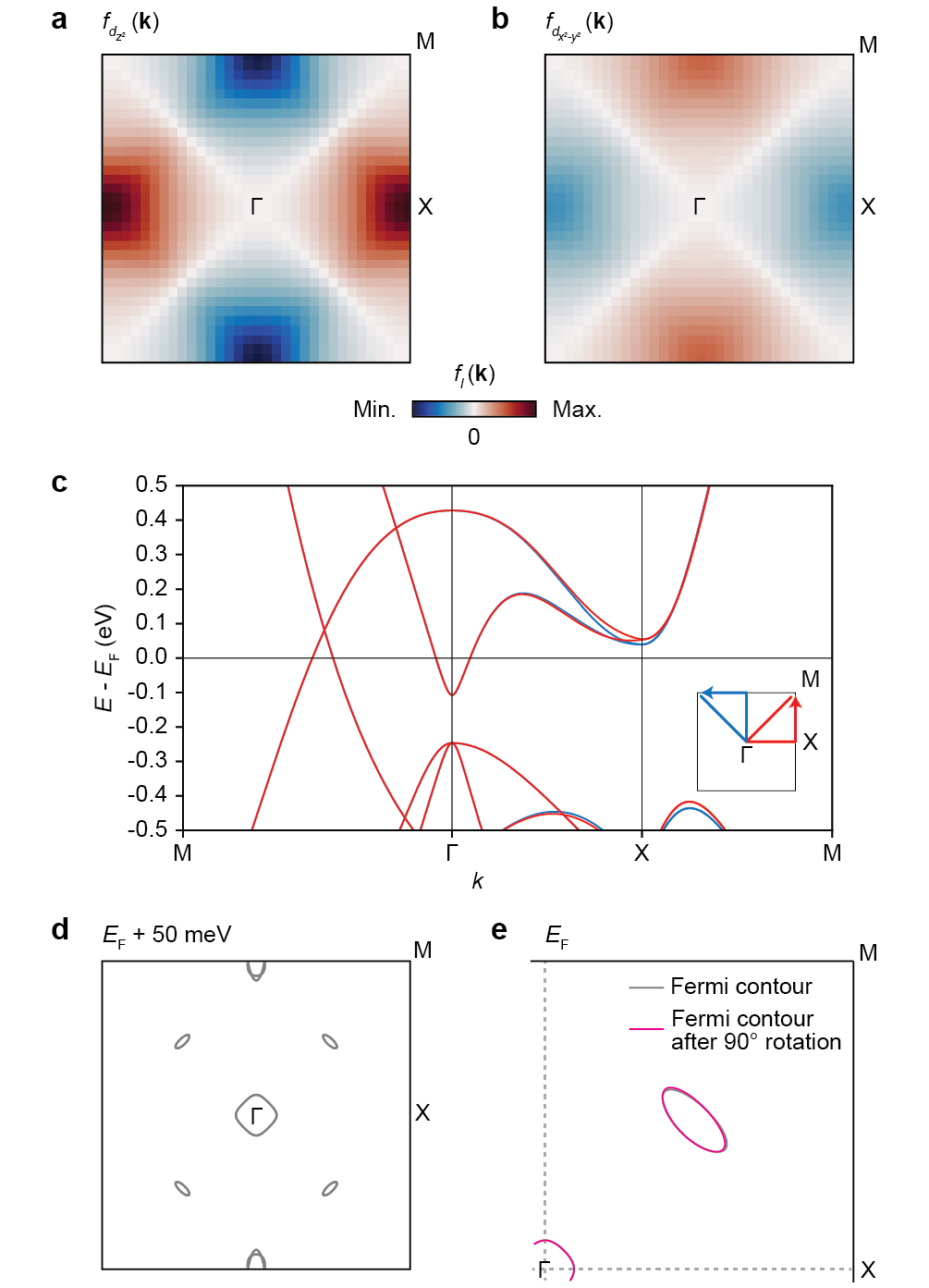}
\caption{\label{fig:5} \textbf{Symmetry breaking corrections \textit{via} the density wave equation.} (a) The calculated form factor $f_{d_{z^{2}}}(\mathbf{k})$. (b) The corresponding form factor $f_{d_{x^{2}-y^{2}}}(\mathbf{k})$. (c) Band structures after corrections, plotted along two paths through the BZ that capture the rotational symmetry breaking (see inset). The corrected curves are shown in red and blue, corresponding to the red and blue paths depicted in the inset. They exhibit a significant difference in energy near the $\mathrm{X}$-point. (d) A constant-energy band contour plot at $E_{\mathrm{F}}$ + 50~meV, above the bottom of one set of electron-like pockets, but below the bottom of the other. (e) The effect of $d$-form factor corrections on the Dirac cone. The constant energy contour in the positive quadrant of the BZ, at $E_{\mathrm{F}}$, showing the contour of a Dirac cone (gray). A contour from another Dirac cone, rotated by 90$^{\circ}$ into the positive quadrant, is also shown (magenta).}
\end{figure}

Figure 5 shows results of numerical calculations of $d$-form factor energy corrections $f_{l}(\mathbf{k})$. The full details of the calculations are described in the Methods section. Figs. 5(a) and 5(b) show plots of the form factors calculated for the two dominant orbitals of the electron-like band near X, namely $d_{z^{2}}$ and $d_{x^{2}-y^{2}}$ [recall Fig. 2(a) and (c)]. The strength of each $f_{l}(\mathbf{k})$ is plotted with a common color scale. Although $f_{d_{z^{2}}}(\mathbf{k})$ and $f_{d_{x^{2}-y^{2}}}(\mathbf{k})$ are related by a sign change and therefore partially cancel each other, the former is a factor of $\sim$2 stronger than the latter and a significant $d$-form factor correction remains. Note that an inverse Fourier transform of the $d$-form factor results in a $B_{2g}$ bond-order pattern in real-space, similar to that illustrated in Fig. 2(g). Figure 5(c) shows the resulting band structure after applying the correction to the band energies, sampled along two paths in the BZ whose ${\Gamma \mathrm{X}}$ segments run perpendicular to each other, as shown in the inset. The blue and red curves show negligible difference except in the region near the $\mathrm{X}$-point. This difference serves to explain the observations of rotational symmetry breaking in QPI and in high-resolution spectroscopic imaging described in the sections above. Figure 5(d) shows the constant-energy contours of the corrected band structure at the energy of 50~meV, roughly the energy at which nematicity was observed. 
In Fig. 5(e) we plot one quadrant of the BZ showing the Fermi contour of a Dirac cone, and that of another Dirac cone rotated by 90$^{\circ}$ from an adjacent quadrant. This visualizes the effect of the $d$-form factor correction on the Dirac cones, which is seen to be fairly small. This is because the $\Gamma \mathrm{M}$ lines on which the Dirac cones are centered are the nodes of the form factor, as shown in Figs. 5(a) and 5(b). 

A discussion of the energy savings originating from the emergence of $B_{2g}$ bond order is given in the Supplemental Information.

\section*{Conclusion}

To summarize the results of this work, we have characterized the cleaved (001) surface of BaNiS$_{2}$ using STM, where we can access the electronic properties of the uppermost Ni square net. Through tunneling spectroscopy (including Landau level spectroscopy), we determine the approximate energy of the Dirac nodal-line to be $E_{\mathrm{D}} \approx $ 27~meV. We estimate the Fermi velocities on the inner and outer arcs of each Dirac cone, along the $\overline{\Gamma \mathrm{M}}$ line, as $v_{k,{\mathrm{inner}}}$ = 3.09$\times$10$^{5}$~m\,s$^{-1}$ and $v_{k,{\mathrm{outer}}}$ = -2.98$\times$10$^{5}$~m\,s$^{-1}$, respectively. This is in fairly good agreement with previous reports \cite{Nilforoushan2021}.

High-resolution conductance imaging and QPI observations both reveal a lowering of symmetry, from the $C_{4}$ symmetry expected for a material belonging to the $P4/nmm$ space group, to $C_{2}$. The $C_{2}$ configuration consists of a pair of striped patterns appearing at around 60~meV above $E_{\mathrm{F}}$, related to each other by a 90$^{\circ}$ rotation and $\sim$12~meV energy splitting, and oriented along the Ni bond chains. This symmetry-breaking phenomenon appears to respect the translational symmetries of the surface lattice but break its rotational symmetry. Hence it constitutes an example of nematic order.

A plausible explanation for the observed nematicity is the $d$-form factor modification, $f_{l}(\mathbf{k})$, to the hopping integrals on the Ni square net, characterized within the DW equation framework as resulting from interference between spin fluctuations. Because the Dirac points lie on the nodes of $f_{l}(\mathbf{k})$, they are almost unaffected by the nematicity. An additional unusual feature of nematicity in BaNiS$_{2}$ is that it is present even in a semimetal, having very low density-of-states at $E_{\mathrm{F}}$. This is in sharp contrast with mechanisms such as the Peierls-type instability or the band Jahn-Teller effect, which are usually promoted by a large density-of-states at $E_{\mathrm{F}}$ and a correspondingly large available energy saving upon reconfiguration of the Fermi surface.

From the above findings we conclude that, in BaNiS$_{2}$, topological Dirac nodal-lines and symmetry-breaking electronic correlations coexist without the former being significantly perturbed by the latter. However, looking ahead, the investigation of the BaCo$_{1-x}$Ni$_{x}$S$_{2}$ solid solution will be of interest because substitution with Co should be expected to increase the influence of electronic correlations on the electronic behavior. Ultimately, this may allow us to observe the fate of topologically protected Dirac cones as the system is driven toward a Mott transition.

Overall, as well as adding to the understanding of the nearly ideal Dirac nodal-line electronic structure of BaNiS$_{2}$, the observations presented here also advance the general understanding of electronic nematic states in correlated transition metal square lattice systems.

\section*{Materials and Methods}

\subsection*{STM measurements}
Single crystals of BaNiS$_{2}$ were synthesized as described previously \cite{Shamoto1995, Shamoto1996}. Samples were prepared for measurement by cleaving in an ultra-high vacuum chamber ($P \sim 10^{-10}$~Torr) on a cleaving stage held at about 77~K. They were then quickly inserted into a modified Unisoku 1300 low-temperature STM system held at 1.5~K \cite{Hanaguri2006}, equipped with a home-made head similar to that described previously \cite{Machida2018}. Scanning tips were formed by electrochemically etching tungsten wire, and characterized and conditioned using a field ion microscope and mild indentation at a clean Cu(111) surface. To measure differential conductance $\frac{\textrm{d}I}{\textrm{d}V}$, a lock-in technique was used, with modulation frequency $f_{\textrm{mod}}$ = 617.3~Hz. The modulation amplitude $V_{\textrm{mod}}$ used for each measurement, where relevant, is specified in the respective figure caption.

\subsection*{STM data processing}
To prepare $L(\mathbf{r})$ image data for fast Fourier transformation, one of two methods was used. For all data shown in Fig. 2, a Hann window was applied before transformation. For all imaging of QPI patterns in Figs. 3 and 5, Moisan's \textit{`periodic plus smooth image decomposition'} method was used \cite{Moisan2011}. This avoids the introduction by the Hann window of periodic envelope artifacts in the $q$ domain [visible in Fig. S1(b) of the Supplemental Information].

\subsection*{Density functional theory and tight binding model calculations}
To model QPI patterns, we first performed density functional theory (DFT) calculations for bulk BaNiS$_2$ using the Perdew-Burke-Ernzerhof exchange-correlation functional \cite{Perdew1996}, as implemented in the WIEN2k package \cite{Balaha2013}. We also performed additional calculations using a modified Becke-Johnson exchange potential \cite{Tran2009} to account for the many-body correlation effects not treatable within DFT. These calculations are referred to as correlation-corrected DFT, or in short CC-DFT. For both sets of calculations, we sampled the bulk BZ using a $20\times20\times10$ $k$-mesh and muffin-tin $R_{\mathrm{MT}}$ radius for all atoms such that its product with the maximum modulus of reciprocal space $K_\textrm{max.}$ becomes $R_{\mathrm{MT}}K_\textrm{max.}=7.0$. The bulk DFT and CC-DFT calculations were then downfolded into a 20-band tight-binding model using maximally localised Wannier functions \cite{Souza2001, Mostofi2008} incorporating Ni-$3d$ orbitals as the projection centers. The tight-binding Hamiltonian was then used to construct a 100-layer supercell stacked along the crystalline $c$-axis, while spanning the surface BZ over a fine $256\times256$ $k$-mesh. 

As the CC-DFT results more successfully reproduced observed electronic structures, most notably the electron-like pocket around the $\overline{\Gamma}$-point in photoemission measurements \cite{Nilforoushan2021}, these were adopted for the comparisons discussed in this work.

\subsection*{Calculations of QPI patterns}
With the projected wave functions of the topmost two Ni layers, calculations of QPI were carried out using the \textit{T}-matrix formalism, in a manner similar to that described previously \cite{Kohsaka2017}. The lifetime broadening was chosen to be 1~meV and a localized scalar scatterer with a scattering potential of 0.1~eV in the unitary limit was employed. We performed a basis transformation from the lattice model to the continuum model using the Wannier function \cite{Choubey2014} constructed from the tight-binding wavefunctions between $\pm$1~eV at a height of 1~nm above the Ni$_{\mathrm{A}}$ plane. Lock-in broadening and normalization according to the integrated density-of-states were included for direct comparison with the experimental results. Simulations were carried out for a scatterer at both Ni$_{\mathrm{A}}$ and Ni$_{\mathrm{B}}$ sites, and summed to simulated a random distribution of impurities.

\subsection*{Calculations of \textit{d}-form factor correction \textit{via} DW equation.}
Based on the 20-band tight binding model derived from the CC-DFT results described above, a 2-D five band tight-binding model was constructed by ignoring inter-layer hopping, ignoring spin, and moving from the `folded' to the `unfolded' BZ. In this way the Ni$_{\mathrm{A}}$ and Ni$_{\mathrm{B}}$ sites are treated identically. This model is equivalent to one using the folded BZ if the spin-orbit interaction and inter-layer hopping are both ignored. (The results are displayed in Fig. 5 after returning to the folded BZ.)

With this model as a starting point, and following previous works, the linearized DW equation \cite{Kawaguchi2017} can then be used to characterize rotational symmetry breaking in the hopping integrals. As in previous treatments, the DW interaction includes the Maki-Thomson, Hartree, and Aslamazov-Larkin terms, the latter of which captures the effects of interference between spin fluctuations, and can contribute a $d$-form factor correction \cite{Onari2016}. The interaction is quantified by the momentum-dependent form factor $f_{\mathbf{q},l}(\mathbf{k})$ where $l$ denotes a particular orbital. With $\mathbf{q}$ = 0, $f_{l}(\mathbf{k})$ can describe a uniform bond order as suggested by observation.

We set $T$ = 40 meV and $J$/$U$ = 0.1, where $J$ is the Hund's coupling and $U$ is the on-site Coulomb interaction.

\section*{Acknowledgements}
We are grateful to T. Machida and A. Gauzzi for helpful discussions. 
This work was supported by JST CREST Grant No. JPMJCR16F2, and by a Grant-in-Aid for Scientific Research on Innovative Areas `Quantum Liquid Crystals' (KAKENHI Grant No. JP19H05824 and No. JP19H05825) from JSPS of Japan. C.J.B. acknowledges support from RIKEN's Programs for Junior Scientists.

\section*{Author contributions}
S. Shamoto synthesized the BaNiS$_{2}$ crystals. C.J.B. performed all STM measurements and related data processing. M. S. B. performed the DFT and tight-binding calculations and Y. K. performed the subsequent \textit{T}-matrix calculations of QPI patterns. Y. Y., S. O. and H. K. proposed and calculated the $d$-form factor band correction. T. H. supervized the project. All authors were involved in interpretation of the results. C. J. B. prepared the manuscript with input from all authors.

\section*{Data availability}
The data that support the findings presented here are available from the corresponding authors upon reasonable request.


\begin{thebibliography}{99}

\bibitem{Keimer2017}
B. Keimer and J. E. Moore,
\textit{The physics of quantum materials.}
Nature Physics \textbf{13}, 1045--1055 (2017).
\url{https://doi.org/10.1038/nphys4302}
%

\bibitem{Li2010}
R. Li, J. Wang, X.-L. Qi, and S.-C. Zhang,
\textit{Dynamical axion field in topological magnetic insulators.}
Nature Physics \textbf{6}, 284--288 (2010).
\url{https://doi.org/10.1038/nphys1534}
%

\bibitem{Cao2018}
Y. Cao, V. Fatemi, S. Fang, K. Watanabe, T. Taniguchi, E. Kaxiras, and P. Jarillo-Herrero,
\textit{Unconventional superconductivity in magic-angle graphene superlattices.}
Nature \textbf{556}, 43--853 (2018).
\url{https://doi.org/10.1038/nature26160}
%

\bibitem{Lawler2010}
M. J. Lawler, K. Fujita, J. Lee, A. R. Schmidt, Y. Kohsaka, C. K. Kim, H. Eisaki, S. Uchida, J. C. Davis, J. P. Sethna and E.-A. Kim.
\textit{Intra-unit cell electronic nematicity of the high-$T_{c}$ copper-oxide pseudogap states.}
Nature \textbf{466}, 347--351 (2010).
\url{https://doi.org/10.1038/nature09169}


\bibitem{Fujita2014}
K. Fujita, M. H. Hamidian, S. D. Edkins, C. K. Kim, Y. Kohsaka, M. Azuma, M. Takano, H. Takagi, H. Eisaki, S. Uchida, A. Allias, M. J. Lawler, E.-A. Kim, S. Sachdev and J. C. S. Davis,
\textit{Direct phase-sensitive identification of a \textit{d}-form factor density wave in underdoped cuprates.}
Proc. Natl. Acad. Sci. \textbf{111}, E3026--E3032 (2014).
\url{https://doi.org/10.1073/pnas.1406297111}


\bibitem{Chuang2010}
T.-M. Chuang, M. P. Allan, J. Lee, Y. Xie, N. Ni, S. L. Bud’ko, G. S. Boebinger, P. C. Canfield and J. C. Davis,
\textit{Nematic Electronic Structure in the ``Parent'' State of the Iron-Based Superconductor Ca(Fe$_{1-x}$Co$_{x}$)$_{2}$As$_{2}$.}
Science \textbf{327}, 181--184 (2010).
\url{https://doi.org/10.1126/science.1181083}


\bibitem{Fernandes2014}
R. M. Fernandes, A. V. Chubukov and J. Schmalian,
\textit{What drives nematic order in iron-based superconductors?}
Nature Physics \textbf{10}, 97--104 (2014).
\url{https://doi.org/10.1038/nphys2877}


\bibitem{Kivelson1998}
S. A. Kivelson, E. Fradkin, and V. J. Emery, 
\textit{Electronic liquid-crystal phases of a doped Mott insulator. }
Nature \textbf{393}, 550–553 (1998).
\url{https://doi.org/10.1038/31177}


\bibitem{Oganesyan2001}
V. Oganesyan, S. A. Kivelson, and E. Fradkin, 
\textit{Quantum theory of a nematic Fermi fluid. }
Phys. Rev. B \textbf{64}, 195109 (2001).


\bibitem{Fradkin2010}
E. Fradkin, S. A. Kivelson, M. J. Lawler, J. P. Eisenstein, and A. P. Mackenzie, 
\textit{Nematic Fermi fluids in condensed matter physics. }
Ann. Rev. Cond. Mat. Phys. \textbf{1}, 153–178 (2010).
\url{https://doi.org/10.1146/annurev-conmatphys-070909-103925}


\bibitem{Schoop2018}
L. M. Schoop, F. Pielnhofer, and B. V. Lotsch,
\textit{Chemical Principles of Topological Semimetals.}
Chem. Mater. \textbf{30}, 3155–3176 (2018).
\url{https://doi.org/10.1021/acs.chemmater.7b05133}


\bibitem{SantosCottin2021}
D. Santos-Cottin, M. Casula, L. de' Medici, F. Le Mardel\'{e}, J. Wyzula, M. Orlita, Y. Klein, A. Gauzzi, A. Akrap, and R. P. S. M Lobo,
\textit{Optical conductivity signatures of open Dirac nodal lines.}
Phys. Rev. B \textbf{104}, L201115 (2021).
\url{https://doi.org/10.1103/PhysRevB.104.L201115}


\bibitem{Nilforoushan2021}
N. Nilforoushan, M. Casula, A. Amaricci, M. Caputo, J. Caillaux, L. Khalil, E. Papalazarou, P. Simon, L. Perfetti, I. Vobornik, P. K. Das, J. Fujii, A. Barinov, D. Santos-Cottin, Y. Klein, M. Fabrizio, A. Gauzzi, and M. Marsi,
\textit{Moving Dirac nodes by chemical substitution.}
Proc. Natl. Acad. Sci. \textbf{118}, e2108617118 (2021).
\url{https://doi.org/10.1073/pnas.2108617118}


\bibitem{Nilforoushan2020}
N. Nilforoushan, M. Casula, M. Caputo, E. Papalazarou, J. Caillaux, Z. Chen, L. Perfetti, A. Amaricci, D. Santos-Cottin, Y. Klein, A. Gauzzi, and M. Marsi,
\textit{Photoinduced renormalization and electronic screening of quasi-two-dimensional Dirac states in BaNiS$_{2}$.}
Phys. Rev. Res. \textbf{2}, 043397 (2020).
\url{https://doi.org/10.1103/PhysRevResearch.2.043397}


\bibitem{Mandrus1997}
D. Mandrus, J. L. Sarrao, B. C. Chakoumakos, J. A. Fernandez-Baca, S. E. Nagler, and B. C. Sales,
\textit{Magnetism in BaCoS$_{2}$.}
J. Appl. Phys. \textbf{81}, 4620 (1997).
\url{https://doi.org/10.1063/1.365182}


\bibitem{Krishnakumar2001}
S. R. Krishnakumar, T. Saha-Dasgupta, N. Shanthi, Priya Mahadevan, and D. D. Sarma
\textit{Electronic structure of and covalency driven metal-insulator transition in BaCo$_{1-x}$Ni$_{x}$S$_{2}$.}
Phys. Rev. B \textbf{63}, 045111 (2001).
\url{https://doi.org/10.1103/PhysRevB.63.045111}


\bibitem{Sato2001}
T. Sato, H. Kumigashira, D. Ionel, T. Takahashi, I. Hase, H. Ding, J. C. Campuzano, and S. Shamoto,
\textit{Evolution of metallic states from the Hubbard band in two-dimensional Mott system BaCo$_{1-x}$Ni$_{x}$S$_{2}$.}
Phys. Rev. B \textbf{64}, 075103 (2001).
\url{https://doi.org/10.1103/PhysRevB.64.075103}


\bibitem{Guguchia2019}
Z. Guguchia, B. A. Frandsen, D. Santos-Cottin, S. C. Cheung, Z. Gong, Q. Sheng, K. Yamakawa, A. M. Hallas, M. N. Wilson, Y. Cai, J. Beare, R. Khasanov, R. De Renzi, G. M. Luke, S. Shamoto, A. Gauzzi, Y. Klein, and Y. J. Uemura,
\textit{Probing the quantum phase transition in Mott insulator BaCoS$_{2}$ tuned by pressure and Ni substitution.}
Phys. Rev. Materials \textbf{3}, 045001 (2019).
\url{https://doi.org/10.1103/PhysRevMaterials.3.045001}


\bibitem{Schueller2020}
E. C. Schueller, K. D. Miller, W. Zhang, J. L. Zuo, J. M. Rondinelli, S. D. Wilson, and R. Seshadri,
\textit{Structural signatures of the insulator-to-metal transition in BaCo$_{1-x}$Ni$_{x}$S$_{2}$.}
Phys. Rev. Materials \textbf{4}, 104401 (2020).
\url{https://doi.org/10.1103/PhysRevMaterials.4.104401}


\bibitem{SantosCottin2016a}
D. Santos-Cottin, A. Gauzzi, M. Verseils, B. Baptiste, G. Feve, V. Freulon, B. Pla\c{c}ais, M. Casula, and Y. Klein,
\textit{Anomalous metallic state in quasi-two-dimensional BaNiS$_{2}$.}
Phys. Rev. B \textbf{93}, 125120 (2016).
\url{https://doi.org/10.1103/PhysRevB.93.125120}


\bibitem{Grey1970}
I. E. Grey and H. Steinfink,
\textit{Crystal structure and properties of barium nickel sulfide, a square-pyramidal nickel(II) compound.}
J. Am. Chem. Soc. \textbf{92}, 17, 5093--5095 (1970).
\url{https://doi.org/10.1021/ja00720a015}


\bibitem{VESTA}
K.  Momma and F.  Izumi,
\textit{VESTA 3 for three-dimensional visualization of crystal, volumetric and morphology data.}
J. Appl. Crystallogr. \textbf{44}, 1272--1276 (2011).
\url{https://doi.org/10.1107/S0021889811038970}


\bibitem{Klein2018}
Y. Klein, M. Casula, D. Santos-Cottin, A. Audouard, D. Vignolles, G. F\`{e}ve, V. Freulon, B. Pla\c{c}ais, M. Verseils, H. Yang, L. Paulatto, and Andrea Gauzzi,
\textit{Importance of nonlocal electron correlation in the BaNiS$_{2}$ semimetal from quantum oscillations studies.}
Phys. Rev. B \textbf{97}, 075140 (2018).
\url{https://doi.org/10.1103/PhysRevB.97.075140}


\bibitem{SantosCottin2016b}
D. Santos-Cottin, M. Casula, G. Lantz, Y. Klein, L. Petaccia, P. Le F\`{e}vre, F. Bertran, E. Papalazarou, M. Marsi, and A. Gauzzi,
\textit{Rashba coupling amplification by a staggered crystal field.}
Nat. Commun. \textbf{7}, 11258 (2016).
\url{https://doi.org/10.1038/ncomms11258}


\bibitem{Slawinska2016}
J. S\l{}awi\'{n}ska, A. Narayan, and S. Picozzi,
\textit{Hidden spin polarization in nonmagnetic centrosymmetric BaNiS$_{2}$ crystal: Signatures from first principles.}
Phys. Rev. B \textbf{94}, 241114 (2016).
\url{https://doi.org/10.1103/PhysRevB.94.241114}


\bibitem{Kohsaka2007}
Y. Kohsaka, C. Taylor, K. Fujita, A. Schmidt, C. Lupien, T. Hanaguri, M. Azuma, M. Takano, H. Eisaki, H. Takagi, S. Uchida, and J. C. Davis,
\textit{An Intrinsic Bond-Centered Electronic Glass with Unidirectional Domains in Underdoped Cuprates.}
Science \textbf{315,} 1380--1385 (2007).
\url{https://doi.org/10.1126/science.1138584}


\bibitem{Fujioka2019}
J. Fujioka, R. Yamada, M. Kawamura, S. Sakai, M. Hirayama, R. Arita, T. Okawa, D. Hashizume, M. Hoshino, and Y. Tokura,
\textit{Strong-correlation induced high-mobility electrons in Dirac semimetal of perovskite oxide.}
Nat. Commun. \textbf{10}, 362 (2019).
\url{https://doi.org/10.1038/s41467-018-08149-y}


\bibitem{Kawaguchi2017}
K. Kawaguchi, Y. Yamakawa, M. Tsuchiizu and H. Kontani,
\textit{Competing Unconventional Charge-Density-Wave States in Cuprate Superconductors: Spin-Fluctuation-Driven Mechanism.}
J. Phys. Soc. Jpn. \textbf{86}, 063707 (2017).
\url{https://doi.org/10.7566/JPSJ.86.063707}


\bibitem{Onari2012}
S. Onari and H. Kontani,
\textit{Self-consistent Vertex Correction Analysis for Iron-based Superconductors: Mechanism of Coulomb Interaction-Driven Orbital Fluctuations.}
Phys. Rev. Lett. \textbf{109}, 137001 (2012).
\url{https://doi.org/10.1103/PhysRevLett.109.137001}


\bibitem{Onari2016}
S. Onari, Y. Yamakawa, and H. Kontani,
\textit{Sign-Reversing Orbital Polarization in the Nematic Phase of FeSe due to the $C_{2}$ Symmetry Breaking in the Self-Energy.}
Phys. Rev. Lett. \textbf{116}, 227001 (2016).
\url{https://doi.org/10.1103/PhysRevLett.116.227001}


\bibitem{Onari2019}
S. Onari and H. Kontani,
\textit{Origin of diverse nematic orders in Fe-based superconductors: rotated nematicity in \textit{A}Fe$_{2}$As$_{2}$ (\textit{A} = Cs, Rb).}
Phys. Rev. B \textbf{100}, 020507(R) (2019).
\url{https://doi.org/10.1103/PhysRevB.100.020507}


\bibitem{Ishida2020}
K. Ishida, M. Tsujii, S. Hosoi, Y. Mizukami, S. Ishida, A. Iyo, H. Eisaki, T. Wolf, K. Grube, H. v. L\:{o}hneysen, R. M. Fernandes, and T. Shibauchi,
\textit{Novel electronic nematicity in heavily hole-doped iron pnictide superconductors.}
Proc. Natl. Acad. Sci. \textbf{117}, 6424 (2020).
\url{https://doi.org/10.1073/pnas.1909172117}


\bibitem{Onari2020}
S. Onari and H. Kontani,
\textit{Hidden antiferronematic order in Fe-based superconductor BaFe$_{2}$As$_{2}$ and NaFeAs above $T_{S}$.}
Phys. Rev. Res. \textbf{2}, 042005(R) (2020).
\url{https://doi.org/10.1103/PhysRevResearch.2.042005}


\bibitem{Tazai2022a}
R. Tazai, Y. Yamakawa, S. Onari, and H. Kontani,
\textit{Mechanism of exotic density-wave and beyond-Migdal unconventional superconductivity in kagome metal AV$_{3}$Sb$_{5}$ (A = K, Rb, Cs).}
Science Advances \textbf{8}, eabl4108 (2022).
\url{https://doi.org/10.1126/sciadv.abl4108}


\bibitem{Tazai2022b}
R. Tazai, S. Matsubara, Y. Yamakawa, S. Onari, and H. Kontani,
\textit{A Rigorous Formalism of Unconventional Symmetry Breaking in Fermi Liquid Theory and Its Application to Nematicity in FeSe.}
arXiv:2205.02280 (2022).
\url{https://doi.org/10.48550/arXiv.2205.02280}


\bibitem{Shamoto1995}
S. Shamoto, S. Tanaka, E. Ueda, and M. Sato, 
\textit{Single crystal growth of BaNiS$_{2}$.}
Journal of Crystal Growth \textbf{154}, 197--201 (1995).
\url{https://doi.org/10.1016/0022-0248(95)00225-1}


\bibitem{Shamoto1996}
S. Shamoto, S. Tanaka, E. Ueda, and M. Sato, 
\textit{Single crystal growth of BaCo$_{1-x}$Ni$_{x}$S$_{2}$.}
Physica C \textbf{263}, 550 (1996).
\url{https://doi.org/10.1016/0921-4534(95)00729-6}


\bibitem{Hanaguri2006}
T. Hanaguri,
\textit{Development of high-field STM and its application to the study on magnetically tuned criticality in Sr$_{3}$Ru$_{2}$O$_{7}$.}
J. Phys. Conf. Ser. \textbf{51}, 514 (2006).
\url{https://doi.org/10.1088/1742-6596/51/1/117}


\bibitem{Machida2018}
T. Machida, Y. Kohsaka, and T. Hanaguri,
\textit{A scanning tunneling microscope for spectroscopic imaging below 90 mK in magnetic fields up to 17.5 T.}
Rev. Sci. Instrum. \textbf{89}, 093707 (2018).
\url{https://doi.org/10.1063/1.5049619}


\bibitem{Moisan2011}
L. Moisan,
\textit{Periodic Plus Smooth Image Decomposition.}
Journal of Mathematical Imaging and Vision \textbf{39}, 161–179 (2011).
\url{https://doi.org/10.1007/s10851-010-0227-1}


\bibitem{Perdew1996}
J. Perdew, K. Burke, and M. Ernzerhof, 
\textit{Generalized Gradient Approximation Made Simple}
Phys. Rev. Lett. \textbf{77}, 3865 (1996).
\url{https://doi.org/10.1103/PhysRevLett.77.3865}


\bibitem{Balaha2013}
P. Balaha et al., WIEN2K package, Version 13.1, 2013.


\bibitem{Tran2009}
F. Tran and P. Blaha, 
\textit{Accurate Band Gaps of Semiconductors and Insulators with a Semilocal Exchange-Correlation Potential.}
Phys. Rev. Lett. \textbf{102}, 226401 (2009).
\url{https://doi.org/10.1103/PhysRevLett.102.226401}


\bibitem{Souza2001}
I. Souza, N. Marzari, and D. Vanderbilt, 
\textit{Maximally localized Wannier functions for entangled energy bands.}
Phys. Rev. B \textbf{65}, 035109 (2001).
\url{https://doi.org/10.1103/PhysRevB.65.035109}


\bibitem{Mostofi2008}
[5] A. A. Mostofi, J. R. Yates, Y.-S. Lee, I. Souza, D. Vanderbilt, and N. Marzari, 
\textit{wannier90: A tool for obtaining maximally-localised Wannier functions.}
Comput. Phys. Commun. \textbf{178}, 685 (2008).
\url{https://doi.org/10.1016/j.cpc.2007.11.016}


\bibitem{Kohsaka2017}
Y. Kohsaka, T. Machida, K. Iwaya, M. Kanou, T. Hanaguri, and T. Sasagawa,
\textit{Spin-orbit scattering visualized in quasiparticle interference.}
Phys. Rev. B \textbf{95}, 115307 (2017).
\url{https://doi.org/10.1103/PhysRevB.95.115307}


\bibitem{Choubey2014}
P. Choubey, T. Berlijn, A. Kreisel, C. Cao, and P. J. Hirschfeld,
\textit{Visualization of atomic-scale phenomena in superconductors: Application to FeSe.}
Phys. Rev. B \textbf{90}, 134520 (2014).
\url{https://doi.org/10.1103/PhysRevB.90.134520}


\end{thebibliography}
\end{document}


\title{Supplemental Information for: \protect\\ Correlation-driven electronic nematicity in the Dirac semimetal BaNiS$_{2}$}

\author{C. J. Butler}
\email{christopher.butler@riken.jp}
\affiliation{RIKEN Center for Emergent Matter Science, 2-1 Hirosawa, Wako, Saitama 351-0198, Japan}

\author{Y. Kohsaka}
\affiliation{RIKEN Center for Emergent Matter Science, 2-1 Hirosawa, Wako, Saitama 351-0198, Japan}

\author{Y. Yamakawa}
\affiliation{Department of Physics, Nagoya University, Furo-cho, Nagoya 464-8602, Japan}

\author{M. S. Bahramy}
\affiliation{Department of Physics \& Astronomy, University of Manchester, Oxford Rd., Manchester M13 9PL, United Kingdom}

\author{S. Onari}
\affiliation{Department of Physics, Nagoya University, Furo-cho, Nagoya 464-8602, Japan}

\author{H. Kontani}
\affiliation{Department of Physics, Nagoya University, Furo-cho, Nagoya 464-8602, Japan}

\author{T. Hanaguri}
\email{hanaguri@riken.jp}
\affiliation{RIKEN Center for Emergent Matter Science, 2-1 Hirosawa, Wako, Saitama 351-0198, Japan}

\author{S. Shamoto}
\affiliation{Neutron Science and Technology Center, Comprehensive Research Organization for Science and Society, Tokai, Ibaraki 319-1106, Japan}
\affiliation{Department of Physics, National Cheng Kung University, Tainan, Taiwan 70101, Republic of China}

\maketitle

\newpage

\section*{Further results from STM topography}

\begin{figure*}[t]
\centering
\includegraphics[scale=1]{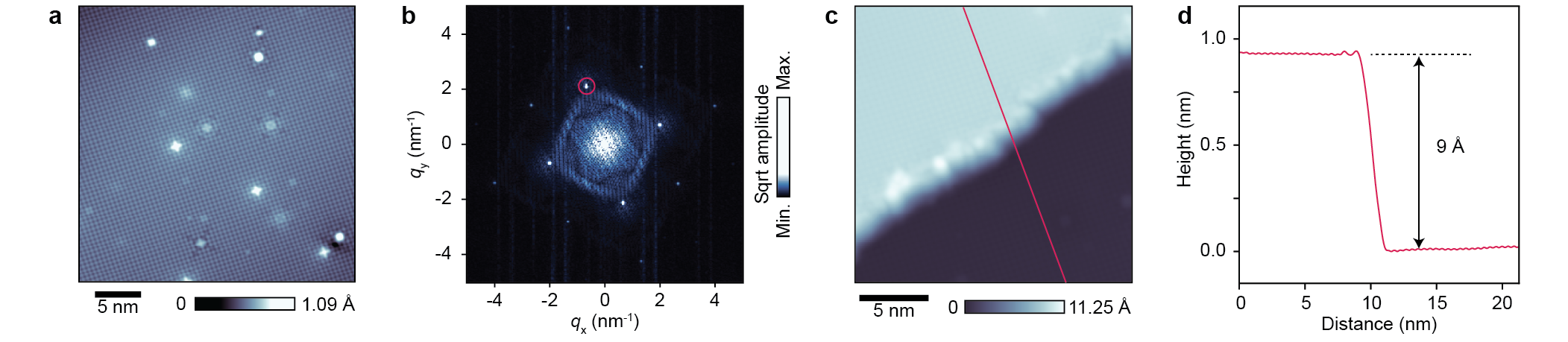}
\caption{
\label{fig:1} \textbf{Additional details from STM topographic images.}
(a) A typical constant-current STM topograph (setpoints \textit{V} = \SI{0.1}{V}, \textit{I} = \SI{100}{pA}), and (b) its Fourier transform. 
The peak circled in red corresponds to a reciprocal lattice vector with a norm of \SI{0.221}{\AA^{-1}} and corresponding periodicity of \SI{4.52}{\AA}, close to the expected lattice parameters \textit{a} and \textit{b}. 
(c) A constant-current topograph of a step between atomic terraces (\textit{V} = \SI{0.1}{V}, \textit{I} = \SI{10}{pA}), and (d) the corresponding height profile along the red line in (c). 
The observed height of \SI{9}{\AA} closely matches the lattice parameter \textit{c}.
}
\end{figure*}

As indicated in Fig. 1(b) of the main manuscript, cleavage can occur between adjacent BaS layers revealing (001) facets.
The STM topography shown in Fig. 1(g) is repeated in Fig. S1(a) below. 
Its Fourier transform $\mathcal{F}[T(\mathbf{r})]$ is shown in Fig. S1(b) and the primary set of reciprocal lattice peaks represent a periodicity of \SI{4.52}{\AA}, close to the reported value for the surface lattice vectors \textit{a} = \textit{b} = \SI{4.43}{\AA} \cite{Grey1970}. 
Although the Ba and apical S are uppermost at the cleaved surface, because the density-of-states at low energies is dominated by Ni orbitals, the atomic corrugation likely represents the Ni square net. 
Moreover, because the observed reciprocal lattice vector indicates only one Ni atom per unit cell is observed, the corrugation therefore represents specifically the uppermost Ni$_{\mathrm{A}}$ square net. 

An additional topography image that includes a step between two atomic terraces is shown in Fig. S1(c), and a height profile, shown in Fig. S1(d), shows an inter-layer separation of \SI{9}{\AA}, close to the lattice parameter of \textit{c} = \SI{8.89}{\AA} \cite{Grey1970}. 

\section*{Landau level spectroscopy}

\begin{figure*}
\centering
\includegraphics[scale=1]{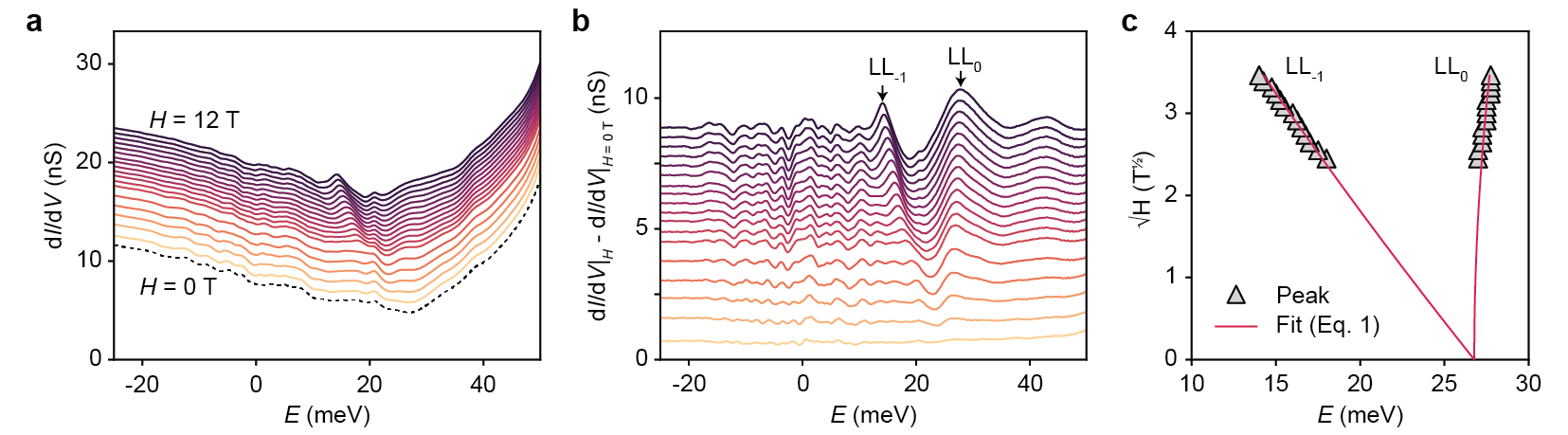}
\caption{\label{fig:3} \textbf{Landau level spectroscopy.} 
(a) $\frac{\textrm{d}I}{\textrm{d}V}(E)$ in the region of $E_{\textrm{F}}$ and $E_{\textrm{D}}$, under magnetic field varying between 0 and \SI{12}{T}. 
The curves are averaged over the field of view marked as a dashed square in Fig. 1(g). 
Each curve is offset vertically by an amount proportional to the applied magnetic field. 
(b) The field-dependence as represented by subtracting the zero-field curve [dashed curve in (a)] from all others. 
Each curve is offset vertically as in (a). 
Anticipating the following result, we label the peak near to $E_{\textrm{D}}$ and the first peak below it as LL$_{0}$ and LL$_{-1}$, respectively. 
(c) Results of fitting the two labeled peaks using Eqn. 1, and using field-dependence curves from 6 to 12~T in increments of 0.5 Tesla.}
\end{figure*}

Under high magnetic fields, in suitably two- or quasi-two-dimensional metals, Landau quantization can be distinguished in conductance spectra measured using STM, in which Landau levels appear as conductance peaks \cite{Cheng2010, Hanaguri2010, Fu2016}. 
Figure S2(a) shows $\frac{\textrm{d}I}{\textrm{d}V}(E)$ spectra acquired under various magnetic fields. 
The magnetic field was applied approximately parallel to the $c$-axis. (Surface tilt was measured to be less than 1$^\circ$.)
Each curve represents the average over the same field of view as for Figs. 1(h) and 1(i) in the main text. 
The magnetic field was increased in increments of \SI{1}{T}, up to \SI{6}{T}, and then in increments of \SI{0.5}{T}, up to the maximum field of \textit{H} = \SI{12}{T} 

A subtle change in the spectra appears in the region between $E$ = 0 and \SI{50}{meV}, and this is clarified upon taking the zero-field conductance curve as a baseline and subtracting it from all other curves, as shown in Fig. S2 (b). 
Two peaks are seen to become prominent approaching \textit{H} = \SI{12}{T}. 
The peak at lower energy shifts significantly with magnetic field, while the peak at higher energy appears to have only a small shift. 
Knowing that under ideal circumstances, the energy of the zeroth Landau level of a topological system should have no field-dependence, we anticipate that the upper of the two peaks can be assigned LL$_{0}$, and the lower of the two can then be assigned as LL$_{-1}$.

Assuming a 2-D Dirac cone, the energy of the $n^{\textrm{th}}$ Landau level can be expressed as

\begin{equation}
E_n = 
E_{\textrm{D}} 
+ sgn(n) \bar{v} \sqrt{2e \hbar(n + \gamma) \mu_{0} H } 
+ g_{\mathrm{LL}} \mu_{\textrm{B}} \mu_{0} H,
\end{equation}

where $\bar{v}$ is the average velocity on the contour of the band from which the Landau level forms, \textit{e} is the electronic charge, $\hbar$ is the reduced Planck constant, $\mu_{0}$ is the vacuum permeability, $g_{\mathrm{LL}}$ is the effective \textit{g}-factor, which we assume to be common to all Landau levels. 
The phase factor $\gamma$ is given by

\begin{equation}
\gamma = 
\frac{1}{2} 
- \frac{\Phi_{\textrm{B}}}{2 \pi}.
\end{equation}

Here $\Phi_{\textrm{B}}$ is the Berry phase accumulated around the Landau orbit. This signifies the topological character of the system and takes the value $\pi$ if the orbit encloses a non-trivial Dirac node. 
In the case that $\gamma$ = 0, and if the Zeeman term is ignored, the energy of the zeroth Landau level (LL$_{0}$) remains at $E_{\textrm{D}}$ independent of magnetic field. 
With the Zeeman term considered, LL$_{0}$ can shift away from $E_{\textrm{D}}$ with increasing magnetic field given a finite \textit{g}-factor $g_{\textrm{LL}}$. 
Using this framework, and adopting the assumption that $\gamma$ = 0, we estimate $E_{\textrm{D}}$, $\overline{v}$ and the two \textit{g}-factors by fitting to the data as shown in Fig S2(c).
For this procedure, after a two-point (0.5~meV) Gaussian filter is applied to smooth each curve, the Landau level peak energies are initially guessed using $E_{\mathrm{peak}, H} = \underset{E}{\mathrm{argmin}} \left[ \frac{\mathrm{d}^{2}}{\mathrm{d}E^{2}} \left( \frac{\textrm{d}I}{\textrm{d}V}_{H}(E)-\frac{\textrm{d}I}{\textrm{d}V}_{H = 0 \textrm{T}} (E) \right) \right]$.
The fitting estimation yields $E_{\textrm{D}} \approx \SI{26.8}{meV}$,  $\overline{v} \approx \SI{1.07e5}{m.s^{-1}}$, and $g_{\mathrm{LL}} \approx 1.32$.

To summarize, the observation of a field-independent Landau level implies the topologically non-trivial nature of the Dirac cones in BaNiS$_{2}$.

\section*{Nematicity in a step-terrace morphology}

\begin{figure*}
\centering
\includegraphics[scale=1]{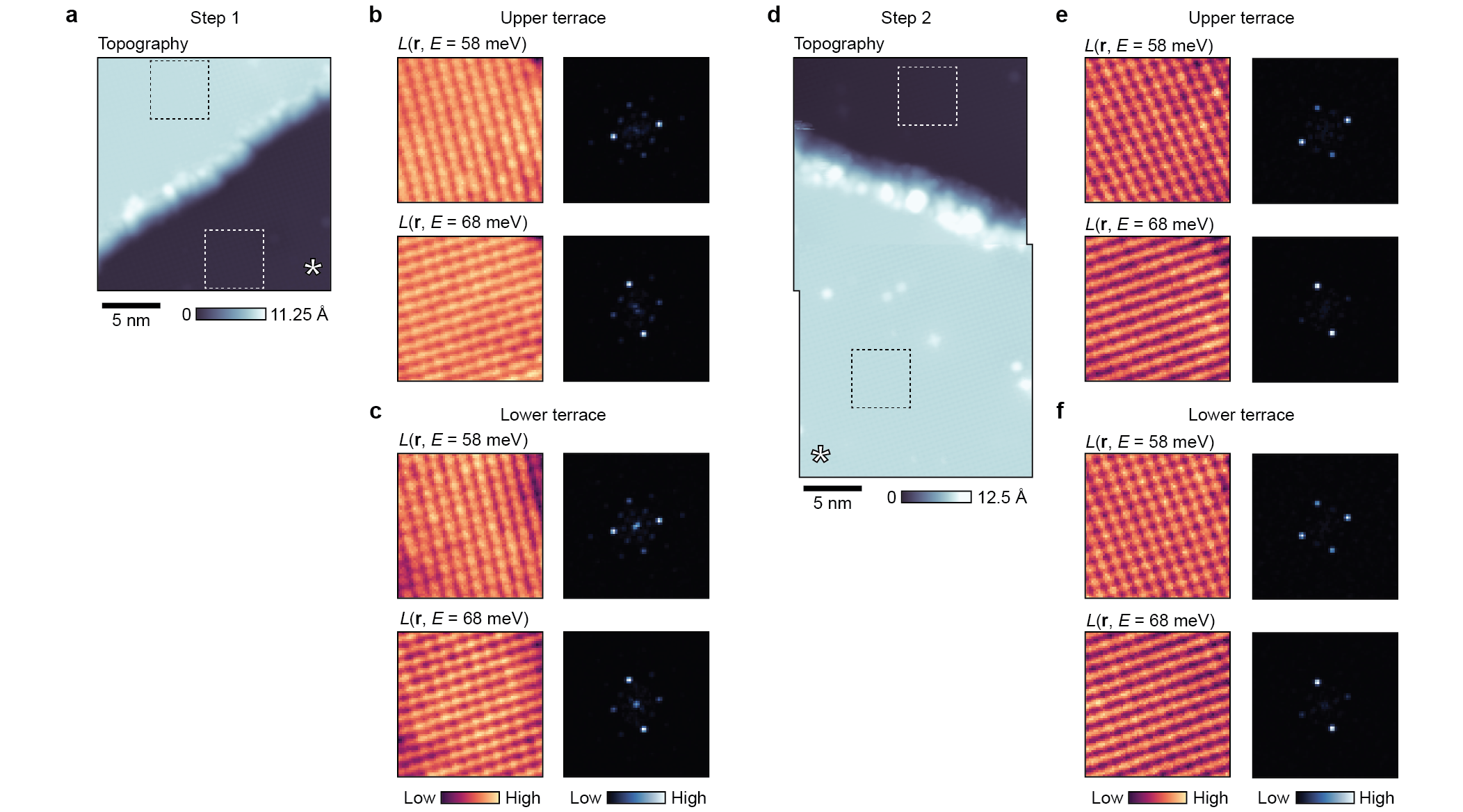}
\caption{\label{fig:3} \textbf{Nematicity in three adjacent terraces.}
(a) Topography showing a single unit-cell high step (setpoints \textit{V} = \SI{0.1}{V}, \textit{I} = \SI{100}{pA}). 
(b) Images of $L(\mathbf{r})$ and their Fourier transforms $L_{q}(\mathbf{q})$ at $E = \SI{58}{mV}$ and $\SI{68}{mV}$. 
(c) The corresponding images acquired on the lower terrace. 
(d) Topography showing another single unit-cell high step. 
The higher terrace (at the bottom of the image) is the same as that shown in the lower right of (a), as indicated by the asterisk symbols. 
(e) and (f) The corresponding $L(\mathbf{r})$ and $L_{q}(\mathbf{q})$ images at $E = \SI{58}{mV}$ and $\SI{68}{mV}$. 
Here nematicity manifests as a differing intensity between the $L_{q}(\mathbf{q})$ peaks corresponding to the reciprocal lattice vectors $\mathbf{G}_{a}$ and $\mathbf{G}_{b}$. 
The orientation of the striped patterns is the same for all three terraces.
}
\end{figure*}

Figure S3 shows measurements across a step-terrace morphology found at a BaNiS$_{2}$ surface. 
Two steps were found, separating three adjacent terraces [the lower terrace, seen at the bottom of Fig. S3(a) is the same one as at the upper terrace seen at the bottom of Fig. S3(d)]. 
Images of $L(\mathbf{r})$ and their Fourier transforms $L_{q}(\mathbf{q})$, acquired with bias separation of \SI{10}{mV}, capture the presence of the nematicity described in the main text. 
The $L(\mathbf{r})$ and $L_{q}(\mathbf{q})$ images acquired on the upper and lower terraces, in the regions marked by the dashed squares in Fig. S3(a), are shown in Figs. S3(b) and S3(c), respectively. 
The corresponding data acquired above and below the second step, in the regions marked by dashed squares in Fig. S3(d), are shown in Figs S3(e) and S5(f), respectively. 

Figures S3(c) and S3(e) represent data collected on the same terrace [as marked by the asterisk in Figs S3(a) and S3(d)], but have a different appearance due to the STM tip becoming contaminated by during the series of measurements. 

Each set of images exhibits nematicity with the same orientation. 
This indicates either that the relation between nematicity in adjacent steps has `ferro'-type order, or more likely, that the the nematicity in adjacent layers is not directly correlated, but rather is coupled to a global condition such as strain.

\section*{Fitting of quasiparticle interference signals}

Figure S4 shows a comparison of linecuts through measured $L_{q}(\mathbf{q}, E)$ data, and full results of fitting that is summarized by the solid red and orange lines shown in Fig. 3(c) of the main text. 
A model consisting of the sum of multiple Lorentzian lineshapes and a constant offset, $c_{q}$, is fitted at each point along the \textit{q} path. This is written as
\begin{equation}
L_{q}^{\mathrm{fit}}(E) = 
\sum_{n=0}^{N} 
\frac{ I_n (\Gamma_n / 2)}{(E - E_n)^2 +(\Gamma_n / 2)^2} 
+ c_{q}
\end{equation}
Here \textit{N} is the number of Lorentzian lineshapes included in the model, which is automatically determined for each \textit{q} point using a peak finding algorithm \cite{scipy.signal.find_peaks} after some smoothing.

\begin{figure*}
\centering
\includegraphics[scale=1]{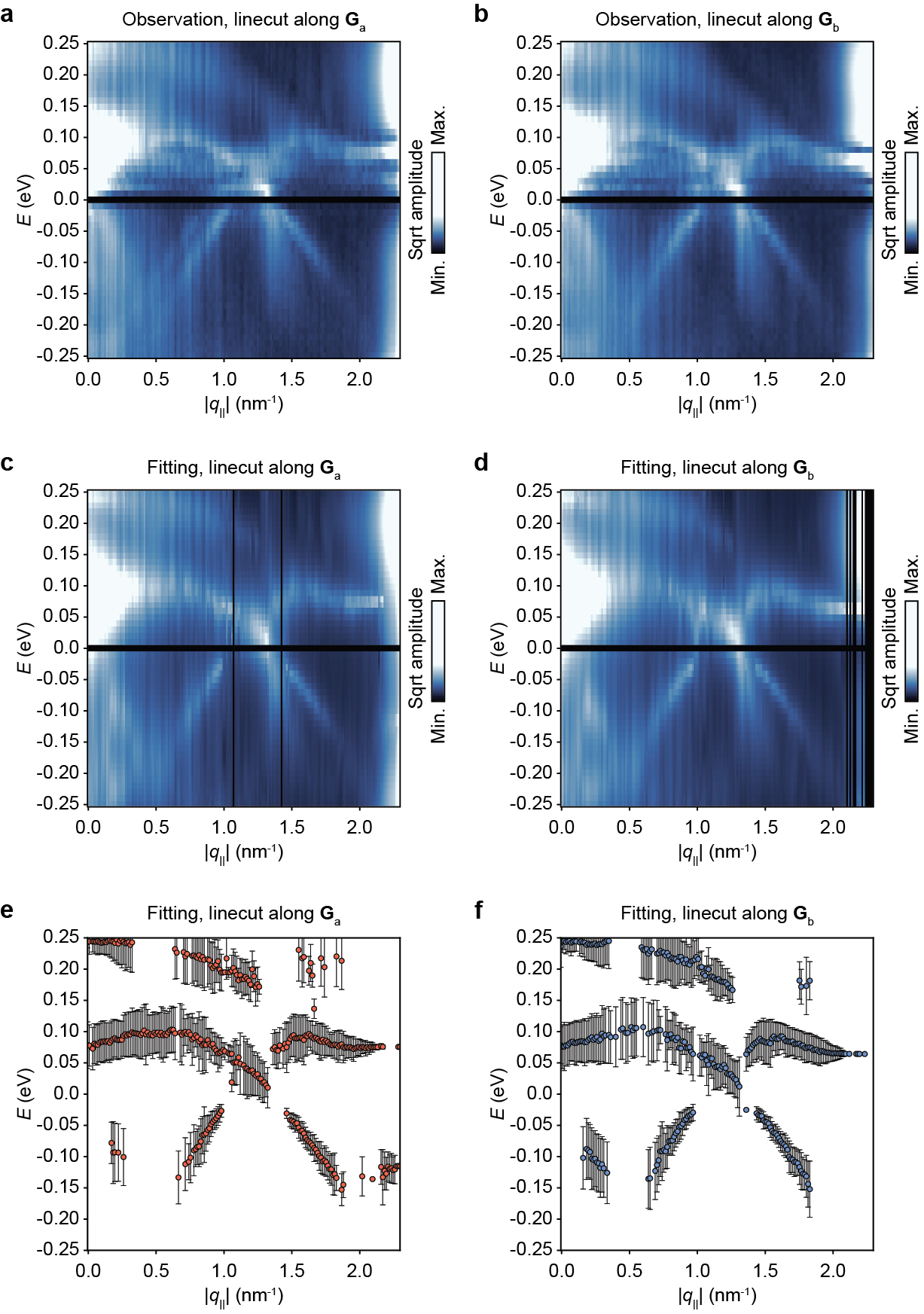}
\caption{\label{fig:4} \textbf{Fitting to $L_{q}(\mathbf{q}, E)$ data.}
(a) and (b) Linecuts through measured $L_{q}(\mathbf{q}, E)$ data, along the $\mathbf{G}_{a}$ and $\mathbf{G}_{b}$ reciprocal lattice vectors. (c) and (d) Corresponding images composed of the fitting results at each $q$ point, with a fitting model described by Eq. 3. Where the fitting procedure failed to converge, a vertical black line is shown. (e) and (f) The same fitting results displayed as points and `error bars' (which in this case denote the Lorentzian broadening $\Gamma$). These plots are superimposed together in Fig. 4(c) of the main text.
}
\end{figure*}

\section*{Quantitative description of Dirac cones derived from QPI results}

As described in the main text, the apparent dispersion of a QPI signal has only an indirect relation to the actual dispersion of the related bands in $k$-space, and for this reason interpretation of QPI observations is generally not straightforward.
In the present case the relation is complicated by the fact that the Dirac cones are not centered at $\mathbf{q} = 0$, and that only scattering between Dirac cones is observed, and not scattering within a single one. 
Nevertheless below we draw a comparison between the observed and calculated features in $q$-space, and also a $k$-space-to-$q$-space comparison of the calculation results, specifically for the QPI signals $\mathbf{q}_{2}$ and $\mathbf{q}_{3}$ shown in Fig. 3 of the main text.
From this we discuss whether the Dirac point energy can be inferred from QPI observations, and also
derive estimates of the Dirac cone velocities parallel to the $\overline{\Gamma \mathrm{M}}$ line.

First, if the observed scattering branches $\mathbf{q}_{2}$ and $\mathbf{q}_{3}$ result from scattering between adjacent Dirac cones as depicted in Fig. 3(d) of the main text, they intersect at the energy where each of the Dirac cones vanishes to a point, so that the shortest and longest possible vectors connecting their contours (namely $\mathbf{q}_{2}$ and $\mathbf{q}_{3}$) become identical. 
The energy at which they intersect ($E = \SI{101}{meV}$) would then seem to be the Dirac point energy. 
However, we find that this is not true. 
The calculated Dirac point energy is in fact about $\SI{60}{meV}$ [see Fig. 1(d) in the main text], meaning the extrapolated point of intersection gives an over-estimate, perhaps due to a finite curvature as the band nears the Dirac point.
This indicates that an extrapolation from the branches $\mathbf{q}_{2}$ and $\mathbf{q}_{3}$ in the observation (which would yield $E_{\mathrm{D}} \approx \SI{47}{meV}$) would similarly over-estimate the actual Dirac point energy. 
For this purpose we fall back on the minimum in the $\frac{\textrm{d}I}{\textrm{d}V}(E)$ curve [Fig. 1(i) of the main text], and also Landau level spectroscopy above, to determine the Dirac point energy. 

As described above, the point of intersection of $\mathbf{q}_{2}$ and $\mathbf{q}_{3}$ represents the $q$-vector connecting two adjacent Dirac points, which we may call $\mathbf{q}_{\textrm{D}-\textrm{D}}$. The assumption that each Dirac point lies on a $\overline{\Gamma \mathrm{M}}$ line in $k$-space leads to $\left| \mathbf{q}_{\textrm{D}-\textrm{D}} \right|^{2} = 2\left| \mathbf{k}_{\textrm{D}}\right|^{2}$ where $\mathbf{k}_{\textrm{D}}$, the wavevector of the Dirac point, is directed from the origin along the $\overline{\Gamma \mathrm{M}}$ line. This results in $\left| \mathbf{k}_{\textrm{D}}\right| \approx \SI{0.884}{nm^{-1}}$.

An approximate value for the Dirac cones' Fermi velocity $v_{\mathrm{F}} = \frac{\textrm{d}E_{\mathrm{F}}}{\textrm{d} \mathbf{k}}$ along a particular axis can be inferred from the apparent velocity $v_{q,2,3}$ of the scattering branches $\mathbf{q}_{2}$ and $\mathbf{q}_{3}$.
We start from the assumption that $v_{q}$ for a generic scattering branch always relates to the velocity $v_{k}$ of a chosen point on the band by a proportionality constant that captures geometric details of the scattering vectors allowed within the band's structure, i.e. $v_{q} = cv_{k}$. 
(As an example, $v_{q} = \frac{1}{2} v_{k}$ in the extremely simple case of scattering across the diameter of an isotropic Dirac cone). 

Next, we compare the calculated and observed velocities $v_{q}^{\mathrm{obs.}} $ and $v_{q}^{\mathrm{calc.}}$ for each branch [see Figs. 3(c) and 3(f) of the main text].
These velocities are related by a factor we call the renormalization factor $R$, such that $v_{q}^{\mathrm{obs.}} = Rv_{q}^{\mathrm{calc.}}$. 
Since in both calculation and in reality, $v_{q} = cv_{k}$ with the same constant $c$ (determined only by geometry), we also posit that the real band velocity is $v_{k} = Rv_{k}^{\mathrm{calc.}}$. 
In this way a renormalization of the $k$-space band structure can reasonably be inferred from the apparent renormalization of $q$-space structures.

The resulting calculated apparent velocities [see Fig. 3(f) of the main text] are $v_{q_{2}}^{\mathrm{calc.}} = \SI{1.41e5}{m.s^{-1}}$ and $v_{q_{3}}^{\mathrm{calc.}} = \SI{-1.18e5}{m.s^{-1}}$, respectively. 
Comparing these with the experimental values leads to $R$ = 0.58 for both inner and outer branches $\mathbf{q}_{2}$ and $\mathbf{q}_{3}$.

Using the same method of fitting as shown in Fig. S4, but for the surface spectral function shown in Fig. 1(d) of the main text, gives velocities for the inner and outer arcs of the Dirac cone, along the $\overline{\Gamma \mathrm{M}}$ line, of $v_{k,{\mathrm{inner}}}^{\mathrm{calc.}} = \SI{5.32e5}{m.s^{-1}}$, and $v_{k,{\mathrm{outer}}}^{\mathrm{calc.}} = \SI{-5.14e5}{m.s^{-1}}$, respectively. 
Finally, using $R$ = 0.58 allows us to infer that the actual band velocities in the sample are $v_{k,{\mathrm{inner}}} = \SI{3.09e5}{m.s^{-1}}$ and $v_{k,{\mathrm{outer}}} = \SI{-2.98e5}{m.s^{-1}}$.
As a comparison, from recently reported angle-resolved photoemission spectroscopy results, we estimate corresponding absolute velocities of $\left| v_{k,{\mathrm{inner}}} \right| \sim \left| v_{k,{\mathrm{outer}}} \right| \approx \SI{2e5}{m.s^{-1}}$ \cite{Nilforoushan2021}.

The band velocities along the $\overline{\Gamma \mathrm{M}}$ line are higher than the average Fermi velocity as determined through Landau level spectroscopy above. 
This is partly because the Dirac cones are elliptical, having the highest velocity along $\overline{\Gamma \mathrm{M}}$ and the lowest velocity perpendicular to this line. 
It may also be because the Fermi contour generally has a lower velocity as the Dirac cone has some curvature, becoming lower in velocity near $E_{\mathrm{D}}$. 
The estimate of velocities based on QPI dispersions relies on energies between \SI{-50} and \SI{-100}{meV}, where the cone has steeper and more linear dispersion.

\section*{Origin of energy savings due to nematic order}

Here, we discuss why an energy gain results from the present finite-energy nematic order.
Although the Dirac nodes are almost unchanged by the $B_{2g}$ form factor, the band dispersion along the antinodal direction is significantly modified.
First, we discuss the change in the free-particle energy due to the form factor $f_l(\k) = f\times {\bar f}_l(\k)$, where ${\bar f}_l(\k)$ is the normalized form factor ($\max_{l,\k}|{\bar f}_l(\k)|=1$).
In the hole representation, it is given by
\begin{eqnarray}
&&\Delta E^f_{\rm free}\equiv E^f_{\rm free}-E^{f=0}_{\rm free}\approx a f^2+b f^4 ,
\label{eqn:DEf}
\\
&&E^f_{\rm free}= \frac1N \sum_{n,\k}(1-f(\e_{n,\k}^f))(-\e_{n,\k}^f) ,
\label{eqn:Ef}
\end{eqnarray}
where $f(\e)$ is the Fermi distribution function, $\e_{n,\k}^f$ is the $n$-th band dispersion with the form factor.
We perform the numerical study at $T\sim0.01$, where Eq. (\ref{eqn:DEf}) is very close to the free energy because $TS$ is small.
In the present model, the relations $a<0$ and $b>0$ hold, and therefore $\Delta E^f_{\rm free}<0$ at a finite $f$.
This energy gain mainly originates from the electron-like dispersion around the X,Y point (Fig. 5).
In fact, the bottom of this dispersion ($E_0\approx 0.05$~eV) is given as $E_{X(Y)}\approx E_0+ (-) c f + d f^2$, where $c\approx1$ and $d>0$.
In fact, $d\approx (E_0-E_0')^{-1}$ due to the second-order perturbation theory,  and $E_0'\approx -0.5$~eV is the energy of the nearest valence band at X,Y points.
Thus, the negative $a$ in Eq. (\ref{eqn:DEf}) is robustly obtained in the present band-structure.

Next, we discuss the total energy with the electron correlation, $\Delta E^f$.
According to Ref. \cite{Tazai2022b}, it is simply given as
\begin{eqnarray}
\Delta E^f \approx a' f^2+b f^4 ,
\label{eqn:DEf2}
\end{eqnarray}
where $a'=a(1-1/\lambda)$, 
and $\lambda$ is the eigenvalue of the DW equation. 
Here, $\lambda>1(<1)$ below (above) $T_{\rm nem}$.
Because $a<0$ in BaNiS$_2$, $a'$ in Eq. (\ref{eqn:DEf2}) is negative below $T_{\rm nem}$, and the order parameter is given as $|f|=\sqrt{-a'/2b}$.
According to Ref. \cite{Tazai2022b}, the specific heat jump at $T_{\rm nem}$ is $\Delta C/T \sim \alpha N(0)$ with $\alpha=(-{\dot \lambda}T)_{T_{\rm nem}}$,  where ${\dot \lambda}\equiv d\lambda /dT$.
In the BCS superconductor, $\alpha=1$. 
In contrast, $\alpha\ll1$ is expected in the case of the nematic transition at $T=T_{\rm nem}$ \cite{Tazai2022b}.

\section*{Temperature dependence}

A natural next step following on from these results is to experimentally characterize the temperature dependence of the nematic order. 
The energy splitting of \SI{\sim12}{meV} corresponds to a temperature of $\sim$140~K, na\"{i}vely the minimum possible transition temperature for the nematic phase. 
Unfortunately, tunneling spectroscopy observations suffer from an energy broadening that rises faster than $k_{\mathrm{B}}T$ (by a factor of 3.5), meaning that observations similar to those presented in this work probably become impractical before reaching the temperature required to observe the potential melting of the nematic phase.
This means that further investigations using other tools may be necessary to characterize its temperature dependent behavior.